\documentclass[12pt]{article}
\pdfoutput=1
\usepackage[latin1]{inputenc}
\usepackage{amsmath}
\usepackage{amsfonts}
\usepackage{amssymb}
\usepackage{graphicx,psfrag}
\usepackage{geometry}
\usepackage{amssymb,epsfig,subfigure}
\usepackage{hyperref}
\usepackage{enumerate}
\usepackage{slashed}
\usepackage[space]{cite}

\numberwithin{equation}{section}

\newcommand{\mathd}{\mathrm{d}}

\newcommand{\tmop}[1]{\ensuremath{\operatorname{#1}}}
\newcommand{\TeXmacs}{T\kern-.1667em\lower.5ex\hbox{E}\kern-.125emX\kern-.1em\lower.5ex\hbox{\textsc{m\kern-.05ema\kern-.125emc\kern-.05ems}}}

\allowdisplaybreaks

\begin{document}
\thispagestyle{empty}
\setcounter{page}0

\ \\
\vspace{2cm}
\begin{center}
{{\LARGE $\tmop{SL}(2,\mathbb{R})$ covariant conditions for $\mathcal{N} = 1$ flux vacua}}\\ \ \\
\vspace{2.5cm}
Ben Heidenreich\\
\vspace{0.8cm}
{\it Department of Physics, Cornell University, Ithaca, NY 14853 USA} \\
\vspace{0.2cm}
{\tt bjh77@cornell.edu}
\end{center}
\vspace{1 cm}
\noindent 
 Four-dimensional supersymmetric $\mathcal{N} = 1$ vacua of type IIB
  supergravity are elegantly described by generalized complex geometry.
  However, this approach typically obscures the $\tmop{SL} (2, \mathbb{R})$ covariance
  of the underlying theory. We show how to rewrite the pure spinor equations
  of Gra\~{n}a, Minasian, Petrini and Tomasiello (hep-th/0505212) in a manifestly $\tmop{SL} (2, \mathbb{R})$ covariant
  fashion. Solutions to these equations fall into two classes: ``charged''
  solutions, such as those containing D5-branes, and ``chargeless'' solutions,
  such as F-theory solutions in the Sen limit and $\tmop{AdS}_4$ solutions. We derive covariant
  supersymmetry conditions for the chargeless case, allowing general
  $\tmop{SU}(3) \times \tmop{SU}(3)$ structure. The formalism presented here greatly simplifies the study of the ten-dimensional geometry of general supersymmetric compactifications of F-theory.

\newpage

%
%
%
%
%
%



\section{Introduction}\label{sec:intro}

The first step towards understanding the phenomenological implications of
string theory is to understand the four-dimensional vacua of the theory. These
vacua are thought to be numerous~\cite{Douglas:2006es}, though many subtleties remain
to gain a complete understanding of even one example. Progress
has been hindered by the appearance of apparent flat directions (moduli) in
the scalar-field effective potential of supersymmetric vacua. Known solutions
to this problem involve nonperturbative effects~\cite{Kachru:2003aw,Balasubramanian:2005zx,Bobkov:2010rf}.

Supergravity, the low energy limit of string theory, has so far proved to be
an invaluable tool in the pursuit of vacua. In particular, the study of
Calabi-Yau geometry 
has yielded a detailed
understanding of the massless spectra of
unstabilized
$\mathcal{N}= 1$ orientifolds (see e.g.~\cite{Grimm:2004uq}). 
Naturally, one would like to classify all string vacua if possible. While this
is a very difficult problem, significant progress can be made if we restrict
our attention to vacua with a geometric supergravity description, with the
necessary\footnote{In particular, local sources of negative tension (e.g.\ O-planes) are required for a geometric vacuum solution with $\mathcal{N} \le 1$ supersymmetry and a non-negative cosmological constant.} addition of localized brane sources. The geometry of such vacua need
not be Calabi-Yau in any sense, and more general tools are needed.

In the case
of unbroken $\mathcal{N}= 1$ supersymmetry, significant progress in this
direction has been made already using generalized complex
geometry~\cite{Hitchin:2004ut,Gualtieri:2003dx}. Gra\~na, Minasian, Petrini, and Tomasiello~\cite{Grana:2005sn, Grana:2006kf}
have shown that the conditions for unbroken $\mathcal{N}= 1$
supersymmetry in type IIA/B supergravity can be rewritten as a set of
relatively simple algebraic and differential conditions on a pair of
compatible pure spinors. Supersymmetric brane embeddings may also be described
elegantly using generalized calibrations~\cite{Koerber:2005qi,Martucci:2005ht,Martucci:2006ij}. Moreover, a few
explicit examples of solutions of this type are now known in type IIA~\cite{Behrndt:2004km,Behrndt:2004mj,Lust:2004ig,Grana:2006kf,Aldazabal:2007sn,Tomasiello:2007eq,Koerber:2008rx,Andriot:2008va,Petrini:2009ur,Lust:2009mb,Aharony:2010af} and IIB~\cite{Butti:2004pk,Grana:2006kf,Andriot:2008va,Lust:2009mb,Heidenreich:2010ad}.

One serious drawback of the pure-spinor equations of~\cite{Grana:2005sn} is that they obscure
the $\tmop{SL}(2, \mathbb{R})$ invariance of type IIB
supergravity. Though this may seem to be merely an aesthetic problem at first
glance, it becomes a serious obstacle in the presence of nontrivial $\tmop{SL}(2,\mathbb{Z})$ monodromies, such as arise in F-theory
compactifications (see e.g.~\cite{Vafa:1996xn,Denef:2008wq}). Though the pure spinors are invariant under axion shifts,
and transform in a known fashion under orientifold involutions~\cite{Grana:2006kf},
their behavior under more general $\tmop{SL}(2,\mathbb{R})$ transformations is significantly more complicated, and,
we believe, not previously described in the literature.\footnote{Some of the
results presented in this paper were referenced in~\cite{Heidenreich:2010ad} without proof.}\,\,\footnote{In principle, some of these results may be encoded in the far more ambitious approach of exceptional generalized geometry~\cite{Grana:2009im, Aldazabal:2010ef, Grana:2011nb}, though decoding them is far from trivial.}

It is therefore advantageous to restate the $\mathcal{N}= 1$ supersymmetry
conditions in an $\tmop{SL}(2,\mathbb{R})$ covariant fashion,
both as a consistency check, and to better describe the geometry of
stabilized F-theory compactifications. In this paper, we carry out this
computation for a special, yet important, class of ``chargeless'' solutions. In particular, all $\tmop{AdS}_4$ solutions are chargeless, and, we will argue, such solutions encompass all possible supersymmetric deformations of  $\tmop{SU}(3)$
structure F-theory compactifications. We find that the covariant conditions obtained here are easier to work with than the original pure spinor equations, though they are less compactly stated.

While the general conditions for $\mathcal{N}= 1$ supersymmetry are
interesting in their own right, the problem of moduli stabilization provides
additional impetus for studying them. It has been suggested~\cite{Koerber:2007xk} that
gaugino condensation on seven-branes, a key ingredient of moduli stabilization
in several scenarios, sources a generalized complex geometry. An improved
understanding of such geometries (see~\cite{Baumann:2010sx, Heidenreich:2010ad, Dymarsky:2010mf}) could lead to a better understanding of known
scenarios for moduli stabilization, or to additional
scenarios for stabilizing moduli.

In \S\ref{sec:purespinoreqns} we review the algebraic and differential conditions for unbroken $\mathcal{N}=1$ supersymmetry, as laid out by~\cite{Grana:2005sn}.
In \S\ref{sec:SL2Rsetup}, we review the $\tmop{SL}(2,\mathbb{R})$ covariance of type IIB supergravity and rewrite the pure
spinor equations in Einstein frame. We then show that $\mathcal{N}= 1$ solutions fall into two classes, which we call ``charged'' and ``chargeless,''
and review chargeless $\tmop{SU}(3)$ structure, the well-known ``F-theory'' solutions.
In \S\ref{sec:chargelessSU3xSU3}, we show how to characterize the $\tmop{SU}(3) \times \tmop{SU}(3)$ structure of a general chargeless solution
in terms of $\tmop{SL}(2, \mathbb{R})$ covariant forms, and, writing the pure spinors
in terms of these forms, we derive $\tmop{SL}(2,\mathbb{R})$ covariant supersymmetry conditions for these solutions, (\ref{alphathetajeqn} -- \ref{six1eqn}). (The technical details of this computation are presented separately in appendix~\ref{app:derivation}.) In \S\ref{sec:consistency} review the flux equations of motion and show that they follow from the supersymmetry conditions and Bianchi identities. We also show that the flux superpotential proposed in~\cite{Koerber:2007xk} is $\tmop{SL}(2,\mathbb{R})$ invariant. In {\S}\ref{sec:conclusions} we discuss future directions and conclude.

\section{The pure-spinor equations for general $\mathcal{N}= 1$
vacua}\label{sec:purespinoreqns}

The supersymmetry transformations of type IIB supergravity are generated by
two ten-dimensional Majorana-Weyl spinors of the same chirality, which we
denote by $\epsilon^i$, for $i = 1, 2$. For $\mathcal{N}= 1$ solutions, these
should be determined by a single four-dimensional Weyl spinor $\zeta_+$ (the
generator of $\mathcal{N}= 1$ supersymmetry transformations.) The most general
relation compatible with unbroken $\mathcal{N}= 1$ supersymmetry is~\cite{Grana:2005sn}
\begin{equation}
  \epsilon^i = \zeta_+ \otimes \eta_+^i + \zeta_- \otimes \eta_-^i \;,
\end{equation}
where $\eta_+^i$ are a pair of fixed positive-chirality internal spinors and
$\zeta_-$ and $\eta_-^i$ are the Majorana conjugates of $\zeta_+$ and
$\eta_+^i$ respectively.

We construct bispinors $\slashed \Upsilon_+$ and $\slashed \Upsilon_-$ from the spinors
$\eta_+^i$:
\begin{equation}
 \slashed \Upsilon_{\pm} = - 8 i \eta^1_+ \otimes \eta^2_{\pm}{}^{\dag} \; . \label{slashYpm}
\end{equation}
Using the Clifford map, $\slashed \Upsilon_{\pm}$ can be rexpressed as
polyforms $\Upsilon_{\pm}$~\cite{Grana:2006kf}
\begin{equation}
\Upsilon_{\pm} = \sum_p \frac{1}{p!} (\Upsilon_{\pm})^{(p)}_{m_1 \ldots m_p} \mathd y^{m_1}\wedge \ldots \mathd y^{m_p} \longleftrightarrow \slashed \Upsilon_{\pm} = \sum_p \frac{1}{p!} (\Upsilon_{\pm})^{(p)}_{m_1 \ldots m_p} \gamma^{m_1 \ldots m_p}
\end{equation}
where $\gamma^{m_1 \ldots m_p} = \gamma^{[m_1} \ldots \gamma^{m_p]}$ denotes the antisymmetrized product of gamma matrices.

As direct products of spinors, the bispinors $\slashed \Upsilon_{\pm}$ are not generic, and satisfy certain algebraic constraints. We review these constraints in \S\ref{subsec:algebraic}, rewriting them as constraints on the polyforms $\Upsilon_{\pm}$ using the language of $G$-structures. Unbroken supersymmetry imposes additional differential conditions on the polyforms $\Upsilon_{\pm}$, which we review in \S\ref{subsec:differential}. Taken together, the conditions in \S\ref{subsec:algebraic} and \S\ref{subsec:differential} are sufficient for unbroken $\mathcal{N}=1$ supersymmetry, and it is no longer necessary to construct the original spinors $\eta^i_+$ explicitly.

\subsection{Algebraic constraints -- $\tmop{SU}(3)\times\tmop{SU}(3)$ structure} \label{subsec:algebraic}

The algebraic constraints on $\slashed \Upsilon_{\pm}$ can be expressed in terms of pure spinors. A pure spinor is a spinor which is annihilated by one-half of the gamma matrices in the associated Clifford algebra,\footnote{We take the Clifford algebra to be even-dimensional.} the maximum possible number. Spinors of $\tmop{SO}(d)$ for $d\le6$ are always pure, whereas for $d>6$, purity imposes a non-trivial constraint.

Bispinors of $\tmop{SO}(d)$ can be viewed as spinors of $\tmop{SO}(d,d)$, whose associated Clifford algebra consists of the $d$ gamma matrices of $\tmop{SO}(d)$ $\gamma^m$ acting on the left $\Gamma^m_L = \gamma^m\otimes1$ and the same gamma matrices acting on the right combined with the bispinor chirality operator $\Gamma_R^m = (1\otimes\gamma^m) \Gamma$, where $\Gamma= \gamma\otimes\gamma = (-1)^{d/2} (\gamma^1 \ldots \gamma^{d})\otimes (\gamma^1 \ldots \gamma^{d})$. Left-acting and right-acting gamma matrices commute, whereas all gamma matrices anticommute with the chirality operator $\Gamma$. Thus, $\Gamma^m_L$ and $\Gamma^m_R$ satisfy an $\tmop{SO}(d,d)$ Clifford algebra:
\begin{equation}
\{\Gamma^m_L, \Gamma^n_L\} = 2\, \delta^{m n} \;\;,\;\; \{\Gamma^m_R, \Gamma^n_R\} = -2\, \delta^{m n} \;\;,\;\; \{\Gamma^m_L, \Gamma^n_R\} =0
\end{equation}

Viewed as spinors of $\tmop{SO}(6,6)$, the bispinors $\slashed \Upsilon_{\pm}$ are pure; they are annihilated by three left-acting gamma matrices and three right-acting gamma matrices due to the (automatic) purity of the $\eta_{\pm}^i$ in $d=6$. Moreover, the three left-acting gamma matrices  $\Gamma_+^i$ are common annihilators of $\slashed \Upsilon_{\pm}$, and the three right-acting gamma matrices $\Gamma_-^i$ which annihilate $\slashed\Upsilon_+$ also annihilate $\slashed \Upsilon_-^{c} = 8 i \eta^1_- \otimes \eta^2_+{}^{\dag}$, where
$\slashed \Upsilon^{c} = C \slashed \Upsilon^{\star} C^{\dag}$
denotes the charge conjugate of a bispinor and $C$ is the (unitary) $\tmop{SO}(6)$ charge conjugation matrix.\footnote{We use the spinor conventions of~\cite{polchinskivol2}.}

Thus, the bispinors $\slashed \Upsilon_{\pm}$ divide the twelve-dimensional space of bispinor gamma matrices into four (complex) subspaces $\mathcal{V}_+=\{\Gamma_+^i\}$, $\mathcal{V}_-=\{\Gamma_-^i\}$, and $\mathcal{\bar{V}}_{\pm}=\{\Gamma_{\pm}^i{}^{\dag}\}$. The subspaces $\mathcal{V}_+$ and $\mathcal{V}_-$ are positive and negative respectively, in the sense that $\{\Gamma_+^i, \Gamma_+^j{}^{\dag}\}$ is a positive-definite Hermitean matrix, defining an $\tmop{SU}(3)$ Clifford algebra, whereas $\{\Gamma_-^i, \Gamma_-^j{}^{\dag}\}$ is negative-definite.

It turns out that the properties enumerated in the last few paragraphs are sufficient to establish that $\slashed \Upsilon_{\pm}$ take the form (\ref{slashYpm}) for some spinors $\eta^i_+$~\cite{Grana:2005sn}. We say that $\tmop{SO}(6)$ bispinors $\slashed \Upsilon_{\pm}$ are compatible pure spinors if the spaces $\mathcal{V}_{\pm}$ defined above are of dimension three and respectively positive and negative in the sense defined above.\footnote{These conditions are sufficient to establish the purity of $\slashed \Upsilon_{\pm}$, and generalize readily to any even dimension.} Thus, $\slashed \Upsilon_{\pm}$ are compatible pure spinors if and only if they take the form (\ref{slashYpm}). Moreover the $\eta^i_+$ can be reconstructed from $\slashed \Upsilon_{\pm}$, up to the obvious rescaling $\eta_+^1 \to \lambda \eta_+^1$ and $\eta_+^2 \to \lambda^{-1} \eta_+^2$ for real $\lambda$~\cite{Grana:2005sn}.

It is convenient to re-express the requirement of compatible pure spinors in $d=6$ in terms of the polyforms $\Upsilon_{\pm}$ using the language of $G$-structures. In special cases, $\Upsilon_{\pm}$ define $\tmop{SU}(3)$ or $\tmop{SU}(2)$ structures on the tangent bundle. We review the properties of $\tmop{SU}(3)$ and $\tmop{SU}(2)$ structures in \S\ref{subsec:su2su3structures} before applying them to the general case in \S\ref{subsec:su3xsu3structures}, where $\Upsilon_{\pm}$ define an $\tmop{SU}(3)\times\tmop{SU}(3)$ structure on the tangent plus cotangent bundle.

\subsubsection{$\tmop{SU}(3)$ and $\tmop{SU}(2)$ structures}\label{subsec:su2su3structures}

A $G$-structure on a real (complex) $d$-dimensional vector bundle is a subbundle of the associated vector bundle with reduced structure group $G \subset \tmop{GL}(d,\mathbb{R})$ ($G \subset \tmop{GL}(d,\mathbb{C})$). An almost complex structure on a $d$-dimensional manifold defines a $\tmop{GL}(d / 2, \mathbb{C})$ structure on the tangent bundle by
restricting to holomorphic bases. With the addition of a hermitean metric, the
structure group is reduced to $\tmop{U}(d/2)$ by restricting to
orthonormal frames. The structure group may be further reduced to $\tmop{SU}(d/2)$ using a complex non-degenerate ``volume element'', that
is a globally defined decomposable $d/2$ form $\Omega$ satisfying $\Omega \wedge \bar{\Omega} \neq 0$.
In fact, the entire $\tmop{SU}(d/2)$ structure may be specified using $\Omega$ (which specifies the almost
complex structure) and a nondegenerate two-form, $J$, the
K\"{a}hler form, where the two must satisfy the compatibility
condition, $J \wedge \Omega = 0$, and the associated metric must be positive
definite. It is conventional to normalize such that
\begin{equation}
  \frac{1}{n!} J^n = \frac{i^n}{2^n}  \left( - 1 \right)^{n \left( n - 1
  \right) / 2} \Omega \wedge \bar{\Omega} \neq 0 \;,
\end{equation}
which combines the nondegeneracy conditions for $J$ and $\Omega$, where $n = d
/ 2$. The positive-definiteness of the associated metric is equivalent to the
condition:
\begin{equation}
  - i J \left( v, \bar{v} \right) > 0 \;,
\end{equation}
for any nonvanishing holomorphic vector $v$ (i.e. satisfying $v \neq 0
, \iota_{\bar{v}} \Omega = 0$). This last condition is ``topological,''
since the nondegeneracy of $J$ implies that the associated metric is positive
definite everywhere so long as this is true at any one point.

Given an $\tmop{SU}(d/2)$ structure, the defining forms $J$
and $\Omega$ can be uniquely reconstructed.{\footnote{Given a local holomorphic orthonormal frame $\left\{
\theta^1, \ldots, \theta^n \right\}$, $\Omega \equiv \theta^1
\wedge \ldots \wedge \theta^n$ and $J = \frac{i}{2} \delta_{i \bar{j}}
\theta^i \wedge \bar{\theta}^{\bar{j}}$.}}
Moreover, associated to any $\tmop{SU}(d/2)$ structure, we
have a conjugate $\tmop{SU}(d/2)$ structure, defined by $- J$
and $\Omega^{\star}$.

By providing a global decomposition of $\Omega$ into lower-rank forms, we
obtain a reduced structure group. For $d = 6$, the only nontrivial example of
this is an $\tmop{SU}(2)$ structure, where we decompose
\begin{equation}
  \Omega = \Omega_2 \wedge \Theta \; .
\end{equation}
Taking $\Theta$ to be holomorphic and orthonormal with respect to the metric
defined by $\Omega, J$, and $\Omega_2 = \frac{1}{2} \iota_{\bar{\Theta}} \Omega$,
we see that $J_2 \equiv J - J_1$ has rank two, where $J_1 \equiv \frac{i}{2} \Theta
\wedge \bar{\Theta}$. The $\tmop{SU}(2)$ structure, defined by
$\Theta$, $J_2$, and $\Omega_2$, must satisfy{\footnote{The condition
$\Omega_2 \wedge \Omega_2 = 0$ implies that $\Omega_2$ is rank one (viewed as
an antisymmetry matrix,) and therefore decomposable, whereas the conditions $2
J_2^2 = \Omega_2 \wedge \bar{\Omega}_2 \neq 0$ and $J_2 \wedge \Omega_2 = 0$
imply that $J_2$ is rank 2.}}
\begin{equation}
  J_2 \wedge \Omega_2 = \Omega_2 \wedge \Omega_2 = 0 \;  \;, \;  \; \Omega_2
  \wedge \bar{\Omega}_2 = 2 J_2^2 \;  \;, \;  \; J_1 \wedge J_2^2 \neq 0 \;,
  \label{su2conds}
\end{equation}
and the associated metric must be positive definite as necessary and
sufficient conditions for $J_2, \Omega_2, \Theta$ to define an $\tmop{SU}(2)$ structure. This last condition can be restated as the
requirement
\begin{equation}
  - i J_2 \left( v, \bar{v} \right) > 0 \;, 
\end{equation}
for any nonvanishing vector $v$ satisfying $\iota_{\bar{v}} \Omega_2 = \iota_v \Theta
= \iota_{\bar{v}} \Theta = 0$. As before, this requirement is topological, given
the nondegeneracy of $J_2$.

An $\tmop{SU}(2)$ structure induces a natural
decomposition into ``base'' ($J_2 , \Omega_2$) and ``fiber'' ($\Theta,
\bar{\Theta}$) directions; a nonvanishing vector $v$ is said to
``point along the base'' if $\iota_v J_1 = 0$, and along the fiber if $\iota_v J_2 =
0$, and a nonvanishing one-form $\omega$ is said to point along the base if
$\omega \wedge J_2^2 = 0$, and along the fiber if $\omega \wedge J_1 = 0$.

The conditions on an $\tmop{SU}(2)$ structure possess a
remarkable symmetry. To make this manifest, we define a
vector of real two-forms $\Omega^i = \left\{ \tmop{Re} \Omega_2, \tmop{Im}
\Omega_2, J_2 \right\}$ and the real four-form $\omega_4 = J_2^2$. The
$\tmop{SU}(2)$ structure conditions may now be rewritten as\footnote{The algebraic structure is similar to that of hyper-K\"ahler manifolds.}
\begin{equation}
  \Omega^i \wedge \Omega^j = \delta^{i j} \omega_4 \;  \;, \;  \; J_1 \wedge
  \omega_4 \neq 0 \;,
\end{equation}
together with the condition that the associated metric is positive definite,
which can be rewritten as:
\begin{equation}
  - \varepsilon_{i j k} \Omega^i \left( v^j, v^k \right) > 0 \;,
\end{equation}
where $v^i$ is any nonvanishing triplet of real vectors satisfying $\delta_{i
j} \iota_{v^i} \Omega^j = 0$. Thus, given an $\tmop{SU}(2)$ structure $\left(
\Omega^i, \Theta \right)$, we can obtain a new $\tmop{SU}(2)$
structure by performing an (in principle spatially dependent) $\tmop{SO}(3)$ rotation on the $\Omega^i$. Moreover, the induced metric is
invariant under these rotations. The induced $\tmop{SU}(3)$ structure $J=J_1+J_2$, $\Omega = \Omega_2\wedge\Theta$ is \emph{not} invariant. Thus, an $\tmop{SU}(2)$ structure defines many \emph{different} $\tmop{SU}(3)$ structures~\cite{Dall'Agata:2004dk}, depending on the choice of rotation.

\subsubsection{Compatibility and $\tmop{SU}(3) \times \tmop{SU}(3)$ structure}\label{subsec:su3xsu3structures}

We return to the compatibility conditions on $\Upsilon_{\pm}$ using the language of $G$-structures laid out in the previous section.

A bispinor $\slashed \Upsilon$ is pure if and only if the corresponding polyform takes the form
\begin{equation}
  \Upsilon = \Omega_k \wedge e^{B - i J} \;,
\end{equation}
for real two-forms $B$ and $J$, where $\Omega_k$ is a decomposable $k$-form and $k$ is the \emph{type} of the pure spinor. Even (odd) chirality pure spinors have even (odd) type; thus, by (\ref{slashYpm}), $\Upsilon_+$ is even rank and $\Upsilon_-$ is odd rank. It turns out that the only $(\Upsilon_+, \Upsilon_-)$ types consistent with the compatibility conditions are $(0,3)$, $(2,1)$, and $(0,1)$. The types of the pure spinors can change over the compactification manifold, a phenomenon known as ``type-changing.''

The compatibility conditions on $\Upsilon_{\pm}$ can be restated as the requirement that $\Upsilon_{\pm}$ define a local $\tmop{SU}(2)$ structure (types $(0,1)$ or $(2,1)$) or $\tmop{SU}(3)$ structure (type $(0,3)$). If $(0,3)\leftrightarrow(0,1)$ type-changing occurs, then neither $G$-structure is globally defined.

To restate the compatibility conditions in this way, it is helpful to work with normalized pure-spinors. The bispinor norm corresponds to the Mukai pairing:
\begin{equation}
  \left\langle \Psi, \Phi \right\rangle \equiv [\Psi \wedge \text{$\left( - 1
  \right)^{\hat{p}  \left( \hat{p} - 1 \right) / 2}$} \Phi]_{\tmop{top}} \;,
\end{equation}
where $\hat{p}$ measures the rank of a form, $\hat{p} F_p = p F_p$. We have
\begin{equation}
  \langle \Upsilon_+, \bar{\Upsilon}_+ \rangle = \langle \Upsilon_-,
  \bar{\Upsilon}_- \rangle = (- 2 i)^3 f_a f_b \Omega_6 \;,
\end{equation}
where $\Omega_6$ is the volume element of the metric used to define the spinors, and $f_a = \eta_+^{1 \dag} \eta_+^1$, $f_b = \eta_+^{2 \dag} \eta_+^2$. We define normalized polyforms\footnote{The case of vanishing norm, $f_a=0$ or $f_b=0$, is not encompassed by the approach of~\cite{Grana:2005sn}.}
\begin{equation}
\Psi_{\pm} = \frac{1}{\sqrt{f_a f_b}} \Upsilon_{\pm} \label{psipmdef}
\end{equation}
so that $\langle \Psi_+, \bar{\Psi}_+ \rangle = \langle \Psi_-, \bar{\Psi}_- \rangle = (- 2 i)^3 \Omega_6$.

One can show that, wherever $\Psi_{\pm}$ have types $(0, 3)$, they must take the form
\begin{equation}
  \Psi_+ = e^{i \vartheta} e^{- i J} \;  \;, \;  \; \Psi_- = \Omega \;,
  \label{su3struct}
\end{equation}
where $J$ and $\Omega$ define an $\tmop{SU}(3)$ structure and $\vartheta$ is an additional phase factor. By contrast, wherever
$\Psi_{\pm}$ have types $(0, 1)$ or $(2, 1)$, they
must take the form (see e.g.~\cite{Koerber:2010bx}):
\begin{equation}
  \Psi_+ = e^{i \vartheta} e^{- i J_1} \wedge [c_{\varphi} e^{- i J_2} -
  s_{\varphi} \Omega_2] \;  \;, \;  \; \Psi_- = \Theta \wedge [c_{\varphi}
  \Omega_2 + s_{\varphi} e^{- i J_2}] \;, \label{su2struct}
\end{equation}
where $\Theta, J_2, \Omega_2$ define an $\tmop{SU}(2)$
structure, $\varphi$ is the ``spinor angle,'' and we use the shorthands
$c_{\varphi} = \cos \varphi$ and $s_{\varphi} = \sin \varphi$. Though the types of the pure spinors $\Psi_{\pm}$ may vary across the compact
space, either (\ref{su3struct}) or (\ref{su2struct}) must apply at any point; as such, these two
equations are equivalent to the compatibility conditions.

The pure spinors $\Psi_{\pm}$ have types $\left( 2, 1 \right)$ where $\varphi
= \pi / 2$ and types $\left( 0, 3 \right)$ where $\varphi = 0$, though in the
latter case the $\tmop{SU}(2)$ structure need not be well
defined, and only $\Omega = \Theta \wedge \Omega_2$ and $J = J_1 + J_2$ need
be single-valued, defining a (local) $\tmop{SU}(3)$ structure. Otherwise, for generic spinor angles such that $s_{\varphi}, c_{\varphi} \ne 0$, the pure spinors have types $\left(
0, 1 \right)$.

From the perspective of generalized complex geometry, $\Psi_{\pm}$ define an $\tmop{SU}(3)\times\tmop{SU}(3)$ on the tangent plus cotangent bundle~\cite{Gualtieri:2003dx, Koerber:2010bx}. This is not essential to our discussion, though we refer to the compatible pure spinors as an $\tmop{SU}(3)\times\tmop{SU}(3)$ structure for want of a better label.

\subsection{Differential constraints} \label{subsec:differential}

Having stated the algebraic conditions on $\Upsilon_{\pm}$ in the form (\ref{su3struct}, \ref{su2struct}), we now review the differential conditions on $\Upsilon_{\pm}$ and the spinor norms $f_a$ and $f_b$ for unbroken $\mathcal{N}=1$ supersymmetry.

We adopt the supergravity conventions of~\cite{Grana:2005sn,Grana:2006kf} in string-frame. Subsequently, we employ compatible Einstein-frame conventions, which are outlined in \S\ref{subsec:SL2Rcovar}. The string-frame compactification metric takes the form:
\begin{equation}
  \mathd s^2_{\left( S \right)} = e^{2 A^{\left( S \right)} \left( y \right)}
  \mathd s_{\left( 4 \right)}^2 + g^{\left( S \right)}_{m n} \left( y \right)
  \mathd y^m \mathd y^n \;, \label{strframemetric}
\end{equation}
where $A^{\left( S \right)}$ is the string-frame warp factor, $g^{\left( S
\right)}_{m n}$ is the string-frame warped metric, and $\mathd s_{\left( 4
\right)}^2$ is a maximally isotropic four-dimensional metric (either Minkowski
or anti-de-Sitter.) We define the field-strength polyform
\begin{equation}
  F = F_1 + \tilde{F}_3 + \tilde{F}_5^{\left( \tmop{int} \right)} \;,
\end{equation}
where $\tilde{F}_5^{\left( \tmop{int} \right)}$ denotes the internal
components of $\tilde{F}_5 = \left( 1 + \star_{10} \right) \tilde{F}_5^{\left(
\tmop{int} \right)} = \tilde{F}_5^{\left( \tmop{int} \right)} + e^{4 A^{\left(
S \right)}} \Omega_4 \wedge \star_6^{\left( S \right)} \tilde{F}_5^{\left(
\tmop{int} \right)}$,\footnote{To establish sign conventions, the Hodge star
associated with a $D$-dimensional metric $g$ with volume form $\Omega_{(g)}$
is defined by $\star [\mathd x^{m_1} \wedge \ldots \wedge \mathd x^{m_p}] =
\frac{1}{(D - p) !} \hspace{0.25em} \Omega_{(g)}^{m_1 \ldots m_p}{}_{m_{p + 1}
\ldots m_D}  [\mathd x^{m_{p + 1}} \wedge \ldots \wedge \mathd x^{m_D}]$.} so
that $F$ has internal components only. In terms of $F$, the source-free RR
equations of motion and Bianchi identities take the form
\begin{equation}
  \mathd_H F = 0 \;  \;, \;  \; \mathd_H  \left[ e^{4 A^{\left( S \right)}}
  \star_6^{\left( S \right)} F \right] = 0 \;,
\end{equation}
where $\mathd_H F \equiv \mathd F - H \wedge F$.

The Clifford map is frame-dependent; we denote polyforms constructed using the string-frame metric (\ref{strframemetric}) with a superscript, as in $\Upsilon_{\pm}^{(S)}$, and those constructed using the Einstein-frame metric (\ref{einstmetric}) without. Demanding that the supersymmetry variations vanish, one obtains the differential conditions~\cite{Grana:2005sn}:
\begin{align}
  \mathd_H  \left[ e^{2 A^{\left( S \right)} - \phi} \Upsilon_+^{(S)} \right] &=
  - 3 e^{A^{\left( S \right)} - \phi} \tmop{Re} \left[ \bar{\mu} \Upsilon_-^{(S)}
  \right] - e^{2 A^{\left( S \right)} - \phi} \mathd A^{\left( S \right)}
  \wedge \bar{\Upsilon}_+^{(S)} \nonumber\\
  &\hspace{1cm}+ \frac{1}{2} e^{2 A^{\left( S \right)}}  \left[ (f_a + f_b)
  \hat{\star}_6^{\left( S \right)} F - i (f_a - f_b) F \right] \;, 
  \label{dupsilonpl}\\
  \mathd_H  \left[ e^{2 A^{\left( S \right)} - \phi} \Upsilon_-^{(S)} \right] &=
  - 2 i \mu e^{A^{\left( S \right)} - \phi} \tmop{Im} \left( \Upsilon_+^{(S)}
  \right) \;,  \label{dupsilonmn}\\
  \mathd f_a &= f_b \mathd A^{\left( S \right)} \;\; , \;\; \mathd f_b\; =\; f_a
  \mathd A^{\left( S \right)} \;, \label{fafbeqn}
\end{align}
where $\mu$ is related to the vacuum expectation value of the superpotential, $\langle W \rangle =
\mu / \kappa_4^2$ (see~(\ref{superpotentialvev})), so that the cosomological constant is
$\Lambda = - 3 | \mu |^2$,\footnote{In our conventions, the cosmological
constant is one-quarter of the Ricci scalar: $\Lambda =
R_{\left( 4 \right)} / 4$.} and $\hat{\star}_6 F \equiv \left( - 1
\right)^{\hat{p}  \left( \hat{p} - 1 \right) / 2} \star_6 F$.

The differential conditions on the spinor norms (\ref{fafbeqn}) can be
immediately integrated, giving
\begin{equation}
  f_a = k_0 e^{A^{\left( S \right)}} + k_1 e^{- A^{\left( S \right)}} \;  \;,
  \;  \; f_b = k_0 e^{A^{\left( S \right)}} - k_1 e^{- A^{\left( S \right)}}
  \; .
\end{equation}
Since $f_a, f_b \geqslant 0$, $k_0 > 0$, and we can set $k_0 = 1$ by rescaling
the spinors $\eta^i_+$. From (\ref{psipmdef}), we obtain
$\Psi_{\pm}^{\left( S \right)} = \kappa^{- 1} e^{- A^{(S)}} \Upsilon_{\pm}^{(S)}$,
where $\kappa =  \sqrt{1 - k_1^2 e^{- 4 A^{(S)}}}$. Expressed in terms of $\Psi_{\pm}^{(S)}$, the pure spinor equations (\ref{dupsilonpl},
\ref{dupsilonmn}) become
\begin{align}
  \mathd_H  \left[ \kappa e^{4 A^{\left( S \right)} - \phi} \tmop{Re}
  \Psi_+^{\left( S \right)} \right] &= - 3 \kappa e^{3 A^{\left( S \right)}
  - \phi} \tmop{Re} \left[ \bar{\mu} \Psi_-^{\left( S \right)} \right] + e^{4
  A^{\left( S \right)}}  \hat{\star}_6^{\left( S \right)} F \;, 
  \label{RePsiStrPl}\\
  \mathd_H  \left[ \kappa e^{2 A^{\left( S \right)} - \phi} \tmop{Im}
  \Psi_+^{\left( S \right)} \right] &= - k_1 F \;,  \label{ImPsiStrPl}\\
  \mathd_H  \left[ \kappa e^{3 A^{\left( S \right)} - \phi} \Psi_-^{\left( S
  \right)} \right] &= - 2 i \mu \kappa e^{2 A^{\left( S \right)} - \phi}
  \tmop{Im} \Psi_+^{\left( S \right)} \;,  \label{PsiStrMn}
\end{align}
The conditions (\ref{RePsiStrPl}, \ref{ImPsiStrPl}, \ref{PsiStrMn}), together
with the algebraic conditions on $\Psi_{\pm}^{\left( S \right)}$ (\ref{su3struct}, \ref{su2struct}) are necessary and sufficient
conditions for unbroken four-dimensional $\mathcal{N}= 1$ supersymmetry,
except in the degenerate case $\kappa = 0$~\cite{Grana:2005sn}.\footnote{One can incorporate the degenerate case using $\tmop{SL}(2,\mathbb{R})$ covariance~\cite{future}.} The $\tmop{U}(1)_R$
symmetry associated with four-dimensional $\mathcal{N}= 1$ supersymmetry takes
the form
\begin{equation}
  \Psi_-^{\left( S \right)} \rightarrow e^{i \theta} \Psi_-^{\left( S \right)}
  \;  \;, \;  \; \mu \rightarrow e^{i \theta} \mu \;,
\end{equation}
where the superpotential rotates $W \rightarrow e^{i \theta} W$.

The one-form component of (\ref{ImPsiStrPl}) implies that
\begin{equation}
  k_2 \equiv \kappa e^{2 A^{\left( S \right)} - \phi} \tmop{Im} \Psi_{+ \left(
  0 \right)}^{\left( S \right)} + k_1 C_0 \;, \label{k2defStr}
\end{equation}
is a constant, where the subscript denotes the zero-form component. AdS
$\left( \mu \neq 0 \right)$ solutions to these conditions are more restricted
than Minkowski ($\mu = 0$) solutions. In particular, (\ref{PsiStrMn}) implies
$\tmop{Im} \Psi_{+ \left( 0 \right)}^{\left( S \right)} = 0$ for $\mu \neq 0$.
Moreover, applying $\mathd_H$ to (\ref{PsiStrMn}) and imposing the source-free
Bianchi identity $\mathd H = 0$ as well as (\ref{ImPsiStrPl}), we find $\mu
k_1 F = 0$. Thus, for $\mu \neq 0$, either $k_1$ or $F$ must vanish. However, $F = 0$ implies that $A^{\left( S \right)}$ is constant~\cite{Grana:2005sn},
so that $\kappa$ is constant, and we may take $k_1 = 0$ without altering the
supersymmetry conditions. With this caveat, we conclude that AdS solutions
require $k_1 = k_2 = 0$.

\section{The covariant conditions -- setup}\label{sec:SL2Rsetup}

We wish to restate the $\mathcal{N}= 1$ supersymmetry conditions in a way
which makes the $\tmop{SL} (2, \mathbb{R})$ covariance of type
IIB supergravity manifest. In principle, one could do this by repeating the
steps taken by~\cite{Grana:2005sn, Grana:2006kf} in deriving the pure spinor equations starting with a
manifestly $\tmop{SL} (2, \mathbb{R})$ covariant formulation of
type IIB supergravity and maintaining covariance at each step. However, we
find it more convenient to work with the pure spinors equations
(\ref{RePsiStrPl}, \ref{ImPsiStrPl}, \ref{PsiStrMn}). It is then necessary to
guess how the pure spinors $\Psi_{\pm}$ transform under $\tmop{SL}(2,\mathbb{R})$. This guess can then be validated by showing that the
supersymmetry conditions are covariant.

We examine this last inference in detail. Suppose that we misidentify the
transformation of the pure spinors under $\tmop{SL} (2, \mathbb{R})$, yet find that the supersymmetry conditions are covariant. Cancelling
the $\tmop{SL} (2, \mathbb{R})$ transformation of the
supergravity fields using a genuine $\tmop{SL} (2, \mathbb{R})$
transformation, we find an $\tmop{SL} (2, \mathbb{R})$ symmetry
of the pure spinor equations under which all supergravity fields are invariant
but the pure spinors transform nontrivially. A symmetry of this type can only
be an R-symmetry. Thus, we conclude that there exists a homomorphism from
$\tmop{SL} (2, \mathbb{R})$ to the R-symmetry group $G_R$, so
that $G_R$ contains a subgroup $\tmop{SL} (2, \mathbb{R}) / H$,
where $H$ is a proper normal subgroup of $\tmop{SL} (2, \mathbb{R})$.{\footnote{$H$ must be a proper subgroup because the pure spinors
transform nontrivially by assumption.}} The only possibilites are $H = \left\{ 1 \right\}$,
$H = \left\{ 1, - 1 \right\}$, so that $G_R$ must contain an $\tmop{SL}(2, \mathbb{R})$ or $\tmop{PSL}(2, \mathbb{R})$
subgroup. This is obviously impossible for $\mathcal{N}= 1$ vacua, since then
$G_R \cong \tmop{U}(1)$. In fact, it is still impossible for extended
supersymmetry ($\mathcal{N} \geqslant 2$), since $G_R$ is in general a compact
Lie group, whose Lie algebra does not have subalgebras isomorphic to the split
Lie algebra $\mathfrak{s}\mathfrak{l}_2 (\mathbb{R})$. Thus, we
conclude that the $\tmop{SL} (2, \mathbb{R})$ transformation
properties of the pure spinors are uniquely determined by the covariance of
the pure spinor equations.

This argument relies on the full $\tmop{SL} (2, \mathbb{R})$
invariance of type IIB supergravity. While only an $\tmop{SL}(2,\mathbb{Z})$ subgroup is nonanomalous in the quantum theory, the
conditions derived in~\cite{Grana:2005sn} follow from \emph{classical} type IIB
supergravity, and therefore necessarily possess the full $\tmop{SL}(2,\mathbb{R})$ invariance. Thus, our approach is not only valid, but
additionally presents a highly nontrivial consistency check on the pure spinor
equations (\ref{RePsiStrPl}, \ref{ImPsiStrPl}, \ref{PsiStrMn}), which were
derived without reference to $\tmop{SL} (2, \mathbb{R})$
invariance.

Type IIB supergravity has an even slightly larger, $\tmop{SL}_{\pm}(2,\mathbb{R})$ invariance, where negative and positive determinant
transformations are connected by ``charge conjugation,'' a $\mathbb{Z}_2$
symmetry which reverses all RR fields and leaves the NSNS fields invariant.
Charge conjugation acts simply on the pure spinors, taking $\Psi_{\pm}^{\left(
S \right)} \rightarrow - \Psi_{\pm}^{\left( S \right)}$.\footnote{This is similar to the O5/O9 involution, which combines charge conjugation with $-1 \in \tmop{SL}(2,\mathbb{Z})$, but has a more complicated action on the pure spinors.}

In \S\ref{subsec:SL2Rcovar}, we review how the $\tmop{SL}_{\pm}(2, \mathbb{R})$ invariance of type IIB string theory can be made explicit to develop
the notation necessary to write down the covariant supersymmetry conditions. In \S\ref{subsec:purespinorEF}, we rewrite the pure spinor equations in Einstein frame, and in \S\ref{subsec:chargedchargeless}, we show that solutions fall into two classes, ``charged'' and ``chargeless'' solutions. After reviewing chargeless solutions with strict $\tmop{SU}(3)$-structure in \S\ref{subsec:Fthconds}, the often-studied ``F-theory'' solutions with imaginary self-dual $G_3$ flux, we consider general chargeless solutions in \S\ref{sec:chargelessSU3xSU3}.

\subsection{The $\tmop{SL}_{\pm}(2, \mathbb{R})$ covariance of
type IIB supergravity}\label{subsec:SL2Rcovar}

The bosonic low energy effective action for type IIB string theory written in
Einstein frame is{\footnote{We employ the supergravity conventions of~\cite{Heidenreich:2010ad} in Einstein frame.}}
\begin{align}
  S &= \frac{1}{2 \kappa_{10}^2} \int \mathd^{10} x \sqrt{- g^{(10)}} 
  \left[ R_{\left( 10 \right)} - \frac{1}{2}  \left( (\nabla \phi)^2 + e^{-
  \phi} |H_3 |^2 + e^{2 \phi} |F_1 |^2 + e^{\phi} | \tilde{F}_3 |^2 +
  \frac{1}{2} | \tilde{F}_5 |^2 \right) \right] \nonumber\\
  &\hspace{1cm} - \frac{1}{4 \kappa_{10}^2} \int C_4 \wedge H_3 \wedge F_3 \;, 
\end{align}
where $|F_p |^2 \equiv \frac{1}{p!} F^{M_1 \ldots M_p} F^{\star}_{M_1 \ldots
M_p} = F_p \cdot F_p^{\star}$, $H_3 = \mathd B_2$, $F_p = \mathd C_{p - 1}$,
\begin{equation}
  \tilde{F}_3 = F_3 - C_0 H_3 \;\;, \;\; \tilde{F}_5 = F_5
  - \frac{1}{2} C_2 \wedge H_3 + \frac{1}{2} B_2 \wedge F_3 \;,
\end{equation}
and the equations of motion must be suplemented by the self-duality
constraint, $\tilde{F}_5 = \star_{10} \tilde{F}_5$. The bosonic fields may be
arranged into singlets, doublets, and triplets as follows:
\begin{equation}
  C_2^i = \left(\begin{array}{c}
    C_2\\
    B_2
  \end{array}\right) \;  \;, \;  \; F_3^i = \left(\begin{array}{c}
    F_3\\
    H_3
  \end{array}\right) \;  \;, \;  \; \phi^{i j} = \left(\begin{array}{cc}
    C_0^2 e^{\phi} + e^{- \phi} & C_0 e^{\phi}\\
    C_0 e^{\phi} & e^{\phi}
  \end{array}\right) \;,
\end{equation}
where $\phi^{i j}$ only carries two degrees of freedom due to the constraint
$\det \phi^{i j} = 1$, and $g^{\left( 10 \right)}$, $C_4$, and $\tilde{F}_5$
are singlets. The action can then be rewritten as
\begin{align}
  S &= \frac{1}{2 \kappa_{10}^2}  \int \mathd^{10} x \sqrt{- g^{\left( 10
  \right)}}  \left[ R_{\left( 10 \right)} + \frac{1}{4} F_1^{i j} \cdot \left(
  F_1 \right)_{i j} - \frac{1}{2} \phi_{i j} F_3^i \cdot F_3^j - \frac{1}{4} |
  \tilde{F}_5 |^2 \right] \nonumber \\
  &\hspace{1cm} + \frac{\varepsilon_{i j}}{8 \kappa_{10}^2} \int C_4
  \wedge F_3^i \wedge F_3^j \; .
\end{align}
where
\begin{equation}
  F_1^{i j} = \mathd \phi^{i j} \;  \;, \;  \; \tilde{F}_5 = \mathd C_4 -
  \frac{1}{2} \varepsilon_{i j} C_2^i \wedge F_3^j \;  \;, \;  \;
  \varepsilon_{12} = - \varepsilon_{21} = - \varepsilon^{12} =
  \varepsilon^{21} = + 1 \;,
\end{equation}
and indices are raised and lowered by left multiplication by $\varepsilon_{i
j}$ or $\varepsilon^{i j}$, so that $\left( F_3 \right)_i = \varepsilon_{i j}
F_3^j$. Invariance of the action under global $\Lambda^i_{\; j} \in
\tmop{SL}_{\pm}(2, \mathbb{R})$ transformations is now
manifest, where
\begin{equation}
  F_3^i \rightarrow \Lambda^i_{\; j} F_3^j \;  \;, \;  \; \phi^{i j}
  \rightarrow \Lambda^i_{\; k} \Lambda^j_{\; l} \phi^{k l} \;  \;, \;  \;
  \tilde{F}_5 \rightarrow \left( \det \Lambda \right)  \tilde{F}_5 \; .
\end{equation}
An $\tmop{SL}_{\pm}(2, \mathbb{Z})$ subgroup of this classical
$\tmop{SL}_{\pm}(2, \mathbb{R})$ symmetry of type IIB
supergravity is an exact gauged symmetry of type IIB string theory, as manifested by the presence of branes (seven-branes and orientifold
planes) in the spectrum carrying monodromies of this type. This symmetry group
has a geometric interpretation in F-theory as modular transformations on an
elliptic fibration.

In some contexts, it is convenient to re-express the bosonic fields in complex
combinations. We define $\tau \equiv C_0 + i e^{- \phi} = \tau_1 + i \tau_2$,
so that under $\tmop{SL} (2, \mathbb{R})$ transformations
\begin{equation}
  \tau \rightarrow \frac{a \tau + b}{c \tau + d} \;  \;, \;  \;
  \begin{pmatrix} a&b\\c&d \end{pmatrix} \in \tmop{SL} (2, \mathbb{R}) \; .
\end{equation}
One can then check that the complex doublet $t^i \equiv \frac{1}{\sqrt{\tau_2}}\bigl( \begin{smallmatrix} \tau \\ 1 \end{smallmatrix} \bigr)$
transforms by an additional phase under $\tmop{SL}(2, \mathbb{R})$:
\begin{equation}
  t^i \rightarrow \left( \frac{\left| c \tau + d \right|}{c \tau + d}
  \right) \Lambda^i_{\; j} t^j \;,
\end{equation}
which motivates the definitions of the following complex combinations:
\begin{equation}
  G_1 \equiv \frac{i}{2} t_i t_j F_1^{i j} = \frac{1}{\tau_2} \mathd
  \tau \;  \;, \;  \; \mathcal{A}_2 \equiv t_i C_2^i =
  \frac{1}{\sqrt{\tau_2}}  \left( C_2 - \tau B_2 \right) \;  \;, \;  \; G_3
  \equiv t_i F_3^i = \frac{1}{\sqrt{\tau_2}}  \left( F_3 - \tau H_3 \right)
  \;,
\end{equation}
all of which transform by a phase under $\tmop{SL}(2, \mathbb{R})$ transformations, and which we label as charge $Q = + 1$, $Q = + 1 /
2$, and $Q = + 1 / 2$ respectively, according to the power $2 Q$ of the phase
factor that they transform by. Notably, half-integer charged
quantities change sign under $- 1 \in \tmop{SL} (2, \mathbb{Z})$
transformations, whereas $\tau$ and integer-charged quantities are
invariant.

Since the phase factor is spatially dependent in general, it is necessary to
introduce a covariant derivative:
\begin{equation}
  \mathrm{D} \Omega \equiv \mathd \Omega + \frac{i Q}{\tau_2} \mathd \tau_1
  \wedge \Omega \;  \;, \;  \; \mathrm{D}^2 \Omega = - \frac{Q}{2} G_1 \wedge
  G_1^{\star} \wedge \Omega \;,
\end{equation}
so that $\mathrm{D} \Omega$ carries the same charge as $\Omega$, though the
operator is no longer nilpotent. We also define the nilpotent covariant
derivatives:
\begin{equation}
  \mathrm{D}_{\pm} \Omega = \mathrm{D} \Omega \pm \frac{i}{2} G_1 \wedge
  \Omega^{\star} \;  \;, \;  \;  \mathrm{D}_{\pm}  \tilde{\Omega} = \mathrm{D}
  \tilde{\Omega} \pm \frac{i}{2} G_1^{\star} \wedge \tilde{\Omega}^{\star}
  \;,
\end{equation}
for $\Omega, \tilde{\Omega}$ of charge $+ 1 / 2$ and $- 1 / 2$ respectively.
However, these operators are not $\mathbb{C}$-linear (e.g. $i
\mathrm{D}_{\pm} = \mathrm{D}_{\mp} i$), and so the usual Leibniz rule is not
obeyed in general, though the following identities may be used:
\begin{align}
  \mathrm{D}_{\pm} \left( \Omega_p \wedge F_q \right) &= \mathrm{D}_{\pm}
  \Omega_p \wedge F_q + \left( - 1 \right)^p \Omega_p \wedge \mathd F_q \;, \\
  \mathd \left( \Omega_p \wedge \tilde{\Omega}_q + c.c. \right) &=
  \mathrm{D}_{\pm} \Omega_p \wedge \tilde{\Omega}_q + \left( - 1 \right)^p
  \Omega_p \wedge \mathrm{D}_{\pm}  \tilde{\Omega}_q + c.c. \;, 
\end{align}
for $F_q$ real and neutral, and $\Omega_p$, $\tilde{\Omega}_q$ of charge $+ 1
/ 2$ and $- 1 / 2$ respectively (or vice versa).

Using this notation, the $G_3$ Bianchi identity becomes $\mathrm{D}_- G_3 =
0$, which is solved locally by $G_3 = \mathrm{D}_- \mathcal{A}_2$. The $G_1$
Bianchi identity becomes $\mathrm{D} G_1 = 0$, and the $\tilde{F}_5$ Bianchi
identity and local solution become
\begin{equation}
  \mathd \tilde{F}_5 = \frac{i}{2} G_3 \wedge G_3^{\star} \;  \;, \;  \;
  \tilde{F}_5 = \mathd C_4 + \frac{i}{4} \mathcal{A}_2 \wedge G_3^{\star} +
  c.c. \; .
\end{equation}
Extended $\tmop{SL}_{\pm}(2, \mathbb{R})$ transformations may
be generated by combining $\tmop{SL} (2, \mathbb{R})$
transformations with charge conjugation, i.e.\ $\bigl(\begin{smallmatrix} -1&0\\0&1\end{smallmatrix}\bigr) \in \tmop{SL}_{\pm}(2, \mathbb{R})$, which acts by
negative complex conjugation:
\begin{equation}
  \tau \rightarrow - \tau^{\star} \;  \;, \;  \; G_3 \rightarrow - G_3^{\star}
  \;  \;, \;  \; \tilde{F}_5 \rightarrow - \tilde{F}_5 \;,
\end{equation}
where $\mathrm{D} \Omega$ and $\mathrm{D}_{\pm} \Omega$ once again have the
same transformations properties as $\Omega$.

\subsection{The pure-spinor equations in Einstein
frame}\label{subsec:purespinorEF}

As a first step towards covariantization, we now show how to rewrite the pure
spinor equations (\ref{RePsiStrPl}, \ref{ImPsiStrPl}, \ref{PsiStrMn}) in terms
of the Einstein-frame quantities. We take the following ansatz for the
Einstein-frame metric:
\begin{equation}
  \mathd s_{10}^2 = e^{2 A (y)}\, \mathd s^2_{(4)}
  + e^{- 2 A (y)} \,g_{mn} (y)\, \mathd y^m \mathd
  y^n \;. \label{einstmetric}
\end{equation}
Thus, the warp-factor $A$ and unwarped metric $g$ are related to their
string-frame counterparts by
\begin{equation}
  A^{\left( S \right)} = A + \phi / 4 \;  \;, \;  \; g_{m n}^{\left( S \right)} =
  e^{\phi / 2 - 2 A} g_{m n} \;,
\end{equation}
where $g^{\left( S \right)}$ is the warped metric which appears in
(\ref{strframemetric}).

It is convenient to work with compatible pure spinors whose associated metric
is $g$ rather than $g^{\left( S \right)}$. From (\ref{su3struct},
\ref{su2struct}), we see that this can be accomplished by the rescaling
$\Psi_{\pm}^{\left( S \right)} \equiv e^{\left( \phi / 4 - A \right) \hat{p}} \Psi_{\pm}$,
We also rewrite the Hodge star as
$\hat{\star}_6^{(S)} F = e^{(2 A - \phi / 2)  (3 - \hat{p})}  \hat{\star}_6 F$,
where $\star_6$ is the Hodge star associated with $g$.

Applying these replacements, the pure spinor equations become:
\begin{align}
  \mathd_H  \left[ \kappa e^{(\phi / 4 - A) \hat{p}} e^{4 A}
  \tmop{Re} \Psi_+ \right] &= - 3 \kappa e^{(\phi / 4 - A) (\hat{p}-3)} e^{\phi / 2} \tmop{Re} \left[ \bar{\mu} \Psi_- \right] +
  e^{(2 A - \phi / 2)  (5 - \hat{p})} e^{2 \phi}  \hat{\star}_6 F \;, \label{RePsiPl}\\
  \mathd_H  \left[ \kappa e^{(\phi / 4 - A) (\hat{p}-2)} \tmop{Im} \Psi_+ \right] &= - k_1 F \;,  \label{ImPsiPl} \\
  \mathd_H  \left[ \kappa e^{(\phi / 4 - A) (\hat{p}-3)} e^{\phi / 2} \Psi_- \right] &= - 2 i \mu \kappa 
  e^{(\phi / 4 - A) (\hat{p}-2)} \tmop{Im} \Psi_+ \;,  \label{PsiMn}
\end{align}
where
\begin{equation}
  \kappa = \sqrt{1 - k_1^2 e^{- 4 A - \phi}} \; .
\end{equation}
Depending on the spinor angle, the pure spinors $\Psi_{\pm}$ must either take
the form (\ref{su3struct}) or (\ref{su2struct}).

The supersymmetry conditions imply that $e^{8 A} \star_6 \tilde{F}_5^{\left(
\tmop{int} \right)}$ is closed. Thus, we may take the local ansatz $e^{8 A}
\star_6 \tilde{F}_5^{\left( \tmop{int} \right)} = \mathd \alpha$. Due to the
ten-dimensional self-duality of $\tilde{F}_5$, this implies
\begin{equation}
  \tilde{F}_5 = \left( 1 + \star_{10} \right) \Omega_4 \wedge \mathd \alpha
  \;,
\end{equation}
so that $\alpha$ is related to the external components of $C_4$ via
$C_4^{\left( \tmop{ext} \right)} = \alpha \Omega_4$, where $\Omega_4$ is the volume-form for the external directions. The Minkowski
supersymmetry conditions actually imply that $e^{8 A} \star_6 
\tilde{F}_5^{\left( \tmop{int} \right)}$ is exact, so that $\alpha$ is
globally defined, though this need not be the case for $\mu \neq 0$.

\subsection{Charged and Chargeless solutions} \label{subsec:chargedchargeless}

Solutions to the pure spinor equations fall into two categories, charged
solutions and chargeless solutions, as we now demonstrate.

Consider the one-form component of (\ref{RePsiPl}), the two-form component of
(\ref{PsiMn}), and the three-form component of (\ref{ImPsiPl}):
\begin{align}
  \mathd \left[ \kappa e^{4 A} \tmop{Re} \Psi_+^{\left( 0 \right)} \right] &=
 - 3 \kappa e^{2 A} \tmop{Re} [ \bar{\mu} \Psi_-^{\left( 1 \right)}] +
  \mathd \alpha \;, \\
  \mathd \left[ \kappa e^{2 A} \Psi_-^{\left( 1 \right)} \right] &= - 2 i
  \mu \kappa \tmop{Im} \Psi_+^{\left( 2 \right)} \;, \\
  \mathd \left[ \kappa \tmop{Im} \Psi_+^{\left( 2 \right)} \right] &= k_2
  H_3 - k_1 F_3 \;, 
\end{align}
where we rewrite the last equation using (\ref{k2defStr}) in the form
\begin{equation}
  k_2 \equiv \kappa e^{2 A - \phi / 2} \tmop{Im} \Psi_+^{\left( 0 \right)} +
  k_1 C_0 \;, \label{k2def}
\end{equation}
in order to eliminate $\tmop{Im} \Psi_+^{\left( 0 \right)}$, where $k_2$ is
constant as a result of the one-form component of (\ref{ImPsiPl}). These
equations are consistent with the $\tmop{SL} (2, \mathbb{R})$
invariance of $\kappa \tmop{Re} \Psi_+^{\left( 0 \right)}$, $\kappa
\Psi_-^{\left( 1 \right)}$, and $\kappa \tmop{Im} \Psi_+^{\left( 2 \right)}$,
provided that we identify
\begin{equation}
  Q^i = \begin{pmatrix} k_2\\ k_1 \end{pmatrix} \;,
\end{equation}
as an $\tmop{SL} (2, \mathbb{R})$ doublet. Indeed, these same
forms define calibrations for space-filling, domain-wall, and cosmic-string D3
branes respectively~\cite{Martucci:2005ht, Martucci:2006ij}, and therefore must be $\tmop{SL} (2, \mathbb{R})$ invariant due to the $\tmop{SL}(2,\mathbb{Z})$ invariance of the D3 brane, as classical supergravity
does not distinguish between $\tmop{SL} (2, \mathbb{R})$ and
$\tmop{SL} (2, \mathbb{Z})$.

As a further consistency check on this proposal, note that the pure spinors
always satisfy
$| \Psi_-^{\left( 1 \right)} |^2 / 2 + | \Psi_+^{\left( 0 \right)} |^2 = 1$.\footnote{The one-form $\Theta$ in (\ref{su2struct}) is
normalized so that $\left| \Theta \right|^2 = g^{- 1} \left( \Theta, \bar{\Theta} \right) = 2$.}
Defining $\eta \equiv \kappa \tmop{Re} \Psi_+^{\left( 0 \right)}$ and $\theta
\equiv \kappa \Psi_-^{\left( 1 \right)}$, and using (\ref{k2def}) to eliminate
$\tmop{Im} \Psi_+^{\left( 0 \right)}$, we obtain:
\begin{equation}
  \left| \theta \right|^2 / 2 + \eta^2 = 1 - k_1^2 e^{- 4 A - \phi} - e^{- 4 A
  + \phi}  \left( k_2 - k_1 C_0 \right)^2 = 1 - \left| \chi \right|^2 \;,
  \label{chisqeqn}
\end{equation}
where
\begin{equation}
  \chi \equiv \frac{e^{- 2 A}}{\sqrt{\tau_2}}  (k_2 - \tau k_1) = e^{- 2 A}
  t_i Q^i \;,
\end{equation}
carries charge $+ 1 / 2$. Under these assumptions, (\ref{chisqeqn}) is
manifestly covariant.

As a final consistency check, note that the $- 1 \in \tmop{SL}(2,\mathbb{Z})$ involution of an O3/O7 plane takes the form~\cite{Grana:2006kf}:
\begin{equation}
  \kappa \Psi_+ \rightarrow \left( - 1 \right)^{\hat{p}  \left( \hat{p} - 1
  \right) / 2} \kappa \bar{\Psi}_+ \;  \;, \;  \; \kappa \Psi_- \rightarrow
  \left( - 1 \right)^{\hat{p}  \left( \hat{p} - 1 \right)} \kappa \Psi_- \; .
\end{equation}
Thus, $\kappa \Psi_-^{\left( 3 \right)}$, $\kappa \tmop{Im}
\Psi_+^{\left( 0 \right)}$, $\kappa \tmop{Re} \Psi_+^{\left( 2 \right)}$,
$\kappa \tmop{Im} \Psi_+^{\left( 4 \right)}$, and $\kappa \tmop{Re}
\Psi_+^{\left( 6 \right)}$ change sign under $- 1 \in \tmop{SL}(2,\mathbb{Z})$, whereas the other components are invariant. Therefore, it is consistent to assume that these components consist of sums of half-integer charged terms, whereas the other components consist of sums of neutral and/or integer charged terms.\footnote{One can verify that this is correct by using the known $\tmop{SL}(2,\mathbb{R})$ transformation law for the supersymmetry generators $\epsilon^i$~\cite{future}. I would like to thank P.~Koerber for helpful discussions and correspondence on this point.}
This is consistent with the $\tmop{SL} (2, \mathbb{R})$
invariance of $\kappa \tmop{Re} \Psi_+^{\left( 0 \right)}$, $\kappa
\Psi_-^{\left( 1 \right)}$, and $\kappa \tmop{Im} \Psi_+^{\left( 2 \right)}$.

We refer to solutions with $Q^i = 0$, whether Minkowski or AdS, as
``chargeless,''{\footnote{Solutions of this type were termed ``AdS-like'' in~\cite{Heidenreich:2010ad}.}} and those with $Q^i \neq 0$ as ``charged,'' due to the presence
of a globaly defined $\tmop{SL} (2, \mathbb{R})$ doublet of
constants. Strict $\tmop{SU}(3)$-structure solutions with supersymmetric five-branes (e.g.~\cite{Maldacena:2000yy}) form a well known class of \emph{charged} solutions, whereas strict $\tmop{SU}(3)$-structure solutions with supersymmetric three- and/or seven-branes (e.g.~\cite{Klebanov:1998hh}) form another well-known class of \emph{chargeless} solutions.

As we saw in \S\ref{subsec:differential}, AdS solutions are
always chargeless. Moreover, calibrated space-filling D3 branes can only occur
in chargeless solutions, as space-filling (anti-) D3 branes are calibrated
where $\eta = + 1$ ($\eta = - 1$), which, by (\ref{chisqeqn}), implies $\chi =
0$. Additionally, the presence of a globally defined charge doublet restricts the allowable monodromies to D7 brane and O5 plane monodromies (or $\tmop{SL}(2,\mathbb{Z})$ conjugates of these, depending on the frame); this is inconsistent with the seven-brane configurations found in F-theory setups. For these reasons we focus on chargeless solutions in this paper, for
which, moreover, the supersymmetry conditions take a somewhat simpler form.

Solutions with $\eta = 1$, sometimes referred to as ``F-theory solutions'' since they arise from compactifications of F-theory on Calabi-Yau four-folds in the Sen limit,
form a special well-studied class of examples. We review the supersymmetry
conditions for this case in the next section, before moving on to consider
general chargeless solutions.

\subsection{F-theory solutions}\label{subsec:Fthconds}

By (\ref{chisqeqn}), $\eta^2 = 1$ implies $\chi = 0$ and $\theta = 0$. Thus,
the pure spinors have types $\left( 0, 3 \right)$, and define an $\tmop{SU}(3)$ structure
\begin{equation}
  \Psi_+ = \eta e^{- i J} \;  \;, \;  \; \Psi_- = i \Omega \;,
  \label{Fthpurespinors}
\end{equation}
where $\Omega$ is a decomposable three-form{\footnote{The $i$ in
(\ref{Fthpurespinors}) is purely coventional.}} and $J$ a nondegenerate real
two-form such that
\begin{equation}
  \frac{i}{8} \Omega \wedge \bar{\Omega} = \frac{1}{6} J^3 \neq 0 \;  \;, \; 
  \; J \wedge \Omega = 0 \; . \label{su3conds}
\end{equation}
The choices $\eta = \pm 1$ are related by charge conjugation, under which
$\Psi_{\pm} \rightarrow - \Psi_{\pm}$. We consider the case $\eta = + 1$. The
pure spinor equations (\ref{RePsiPl}, \ref{ImPsiPl}, \ref{PsiMn}) reduce to
\begin{eqnarray}
  e^{4 A} = \alpha  & \;,\; & 0 = H_3 - e^{\phi} \star_6 
  \tilde{F}_3 \;\;,\;\; \mathd \left[ - \frac{1}{2} e^{\phi} J^2 \right] =
  e^{2 \phi} \star F_1 \;,  \label{Ftheqn1}\\
  \mathd J = 0 & \;,\; & 0 = J \wedge H_3 \;\;,\;\; \mathd [e^{\phi / 2}
  \Omega] = 0 \;\;,\;\; 0 = \Omega \wedge H_3 \; . 
  \label{Ftheqn2}
\end{eqnarray}
The conditions involving three-form flux collectively imply that $G_3$ is
primitive with Hodge type $\left( 2, 1 \right)$. Imposing $\mathd J = 0$, the
last equation of (\ref{Ftheqn1}) implies that $\tau$ is holomorphic. These
conditions, together with $e^{4 A} = \alpha$, are manifestly $\tmop{SL}(2, \mathbb{R})$ covariant if the complex structure associated to
$\Omega$ is taken to be $\tmop{SL} (2, \mathbb{R})$ invariant,
which implies that $J$ is also invariant. The third equation of
(\ref{Ftheqn2}) may be rewritten in the covariant form $\mathrm{D} \Omega =
0$, provided that we take $\Omega$ to carry charge $- 1 / 2$ under $\tmop{SL}(2, \mathbb{R})$, where the equivalence of this expression with
(\ref{Ftheqn2}) follows from the holomorphicity of $\tau$. Thus, the
conditions on chargeless $\tmop{SU}(3)$ structure vacua,
commonly known as ``F-theory'' solutions, may be written in the simple
covariant form:
\begin{equation}
  e^{4 A} = \alpha \;  \;, \;  \; G_3 \wedge J = G_3 \wedge \Omega = 0 \;  \;,
  \;  \; \star G_3 = i G_3 \;  \;, \;  \; \mathd J = 0 \;  \;, \; 
  \frac{1}{\tau_2} \mathd \tau \wedge \Omega = 0 \;, \;  \; \mathrm{D} \Omega
  = 0 \;,
\end{equation}
where we take $\Omega$ to carry charge $- 1 / 2$ and $J$ to be
neutral.

The supersymmetry conditions for $\eta = - 1$ are similar:
\begin{equation}
  e^{4 A} = - \alpha \;  \;, \;  \; G_3 \wedge J = G_3 \wedge \Omega = 0 \; 
  \;, \;  \; \star G_3 = - i G_3 \;  \;, \;  \; \mathd J = 0 \;  \;, \;  \;
  \frac{1}{\tau_2} \mathd \tau \wedge \bar{\Omega} = 0 \;  \;, \;  \;
  \mathrm{D} \Omega = 0 \;,
\end{equation}
except that we must now take $\Omega$ to carry charge $+ 1 / 2$, due to the
fact that $\tau$ is now anti-holomorphic. We address this apparent discrepancy
between the cases $\eta = \pm 1$ in the next section.

\section{The chargeless supersymmetry conditions}\label{sec:chargelessSU3xSU3}

Having classified $\mathcal{N}=1$ flux vacua of type IIB supergravity into charged and chargeless backgrounds, we now consider general chargeless solutions. We show how to describe a general chargeless $\tmop{SU}(3) \times \tmop{SU}(3)$ structure in an $\tmop{SL} (2, \mathbb{R})$ covariant fashion in \S\ref{subsec:chargelessSU3xSU3structure},
and then derive covariant versions of the pure spinor equations for these solutions in \S\ref{subsec:covarPSE}.

\subsection{Chargeless $\tmop{SU}(3) \times \tmop{SU}(3)$ structure}\label{subsec:chargelessSU3xSU3structure}

We have shown that for $\eta = 1$, the K\"ahler form $J$ associated to the
$\tmop{SU}(3)$ structure is $\tmop{SL}(2, \mathbb{R})$ invariant, whereas the holomorphic three-form $\Omega$ carries charge
$- 1 / 2$. Deforming away from $\eta = 1$ slightly, the $\tmop{SU}(3)$ structure decomposes into a local $\tmop{SU}(2)$
structure as follows:
\begin{equation}
  J = J_1 + J_2 \;  \;, \;  \; \Omega = \Omega_2 \wedge \Theta \; .
\end{equation}
Since $\theta = \sqrt{1 - \eta^2} \Theta$ by (\ref{su2struct},
\ref{chisqeqn}), we conclude that $\Theta$ is $\tmop{SL}(2, \mathbb{R})$ invariant, and therefore that $J_2$ is invariant and that $\Omega_2$
carries charge $- 1 / 2$. In order to preserve these charge assignments for
arbitary $\eta$, we rewrite the ansatz (\ref{su2struct}) by performing an
$\tmop{SO}(3)$ rotation on the $\Omega^i$:\footnote{Recall that $e^{i \vartheta} = \pm 1$ for a chargeless solution, where the extra sign can be absorbed by redefinitions.}\,\footnote{A similar basis was used in~\cite{Andriot:2008va}.}
\begin{align}
  \Psi_+ &= e^{- i J_1} \wedge \left[ c_{\varphi}  \left( 1 - \frac{1}{2} J_2^2
  \right) + s_{\varphi} \tmop{Im} \Omega_2 - i J_2 \right]  \;, \label{su2chargeless1} \\
  \Psi_- &= \Theta \wedge \left[ s_{\varphi}  \left( 1 - \frac{1}{2} J_2^2
  \right) - c_{\varphi} \tmop{Im} \Omega_2 + i \tmop{Re} \Omega_2 \right] \;, \label{su2chargeless2}
\end{align}
so that $j \equiv - \tmop{Im} \Psi_+^{\left( 2 \right)} = c_{\varphi} J_1 + J_2$ is manifestly invariant.

While (\ref{su2chargeless1}, \ref{su2chargeless2}) completely specify how the pure spinors
transform under $\tmop{SL} (2, \mathbb{R})$ in the chargeless
case, the $\tmop{SU}(2)$ structure forms $\Theta, J_1, J_2$ and
$\Omega_2$ need not be globally defined if $\tmop{SU}(3)$-structure loci
($s_{\varphi} = 0$) are present. Instead, we consider the charge $- 1 / 2$
forms
\begin{equation}
  \omega \equiv s_{\varphi} \Omega_2 \;  \;, \;  \; \beta = \frac{1}{2}  (1 +
  c_{\varphi}) \Omega_2 \wedge \Theta \;  \;, \;  \; \gamma = \frac{1}{2}  (1
  - c_{\varphi}) \Omega_2 \wedge \bar{\Theta} \;,
\end{equation}
in addition to the invariant forms $\eta = c_{\varphi}$, $\theta = s_{\varphi}
\Theta$ and $j = c_{\varphi} J_1 + J_2$ defined previously. All of these forms
are globally defined up to $\tmop{SL}_{\pm}(2, \mathbb{Z})$
monodromies, as they can be extracted from the pure spinors directly. In
particular,
\begin{eqnarray}
  \eta = \tmop{Re} \Psi_+^{\left( 0 \right)} \;\;,\;\; \theta = \Psi_-^{\left( 1
  \right)} \;\;,\;\; j = - \tmop{Im} \Psi_+^{\left( 2 \right)} & \;,\; &
  \tmop{Im} \omega = \tmop{Re} \Psi_+^{\left( 2 \right)} \;\;,\;\; \tmop{Re}
  \omega = \iota_{\tmop{Re} \theta} \tmop{Im} \Psi_-^{\left( 3
  \right)} \;,  \nonumber \\
  \beta = \frac{1}{2 i}  (1 + \eta) \Psi_-^{(3)} - \frac{1}{2 i} \tmop{Im}
  \omega \wedge \theta & \;,\; & \gamma^{\star} = \frac{1}{2 i}  (1 - \eta)
  \Psi_-^{(3)} + \frac{1}{2 i} \tmop{Im} \omega \wedge \theta \;, 
  \label{purespinorstoforms}
\end{eqnarray}
where the inner product is computed using the associated metric. The original
pure spinors can be reconstructed using only these forms:
\begin{equation}
  \Psi_+ = \eta + \tmop{Im} \omega - i j - \frac{1}{2} \eta j^2 - j \wedge j_1
  - i \tmop{Im} \omega \wedge \tilde{j} + \frac{i}{6} j^3 \;  \;, \;  \;
  \Psi_- = \theta + i (\beta + \gamma^{\star}) - \frac{1}{2} j^2 \wedge \theta
  \;, \label{formstopurespinors}
\end{equation}
where $j_1 \equiv \frac{i}{2} \theta \wedge \bar{\theta} = s_{\varphi}^2 J_1$
and $\tilde{j} \equiv \eta j + j_1 = J_1 + c_{\varphi} J_2$.

We refer to the forms $\eta, \theta, j, \omega, \beta$ and $\gamma$ as the
(chargeless) $\tmop{SU}(3) \times \tmop{SU}(3)$
structure henceforward, since they are collectively equivalent to the
chargeless pure spinors by (\ref{purespinorstoforms}) and (\ref{formstopurespinors}). The compatibility and
purity of $\Psi_{\pm}$ impose certain conditions on the $\tmop{SU}(3) \times \tmop{SU}(3)$ structure forms. These are readily
derived by requiring that
\begin{equation}
  \Theta = \frac{1}{\sqrt{1 - \eta^2}} \theta \;  \;, \;  \; J_2 = \frac{1}{1
  - \eta^2} j_2 \;  \;, \;  \; \Omega_2 = \frac{1}{\sqrt{1 - \eta^2}} \omega
  \;, \label{su2extract}
\end{equation}
define an $\tmop{SU}(2)$ structure for $\eta^2 < 1$, where $j_2
\equiv j - \eta \tilde{j} = s_{\varphi}^2 J_2$, and that
\begin{equation}
  \Omega = \beta \;  \;, \;  \; J = j \;, \label{su3extract}
\end{equation}
or
\begin{equation}
  \Omega = \gamma^{\star} \;  \;, \;  \; J = - j \;, \label{antisu3extract}
\end{equation}
define an $\tmop{SU}(3)$ structure for $\eta = \pm 1$
respectively, where $\omega  , \theta$ vanish in the latter
two cases, $\gamma$ vanishes for $\eta = 1$, and $\beta$ vanishes for $\eta =
- 1$. Writing out these conditions using (\ref{su2conds}, \ref{su3conds}) and
simplifying, we find that the $\tmop{SU}(3) \times \tmop{SU}(3)$ structure must satisfy:
\begin{eqnarray}
  \eta^2 \leqslant 1 & , & \frac{1}{2} j_1 \wedge j^2 + \frac{1}{6} \eta j^3 =
  \frac{i}{4 (1 + \eta^2)}  \left[ \beta \wedge \bar{\beta} - \gamma \wedge
  \bar{\gamma} \right] \neq 0 \;,  \label{su3su3cons1}\\
  \left( 1 - \eta \right) \beta = \frac{1}{2} \omega \wedge \theta & , &
  \left( 1 + \eta \right) \gamma = \frac{1}{2} \omega \wedge \bar{\theta} \;, 
  \label{su3su3cons2}\\
  \omega \wedge \omega = 0 & , & j \wedge \omega = \frac{i}{2} \eta \left[
  \theta \wedge \gamma - \bar{\theta} \wedge \beta \right] \;, 
  \label{su3su3cons3}\\
  \frac{1}{2} \omega \wedge \bar{\omega} = j^2 - \tilde{j}^2 & , & j \wedge
  \beta = j \wedge \gamma = 0 \;,  \label{su3su3cons4}
\end{eqnarray}
as well as the requirement that both $\beta$ and $\gamma$ are decomposable and
the topological condition that the associated metric is positive definite.
These conditions are necessary and sufficient to define an $\tmop{SU}(3) \times \tmop{SU}(3)$ structure, and ensure in particular
that $\omega = \theta = 0$ for $\eta = \pm 1$.

Referring to (\ref{purespinorstoforms}), we see
that charge conjugation $\Psi_{\pm} \rightarrow - \Psi_{\pm}$ has the
following action on the $\tmop{SU}(3) \times \tmop{SU}(3)$ structure:
\begin{equation}
  \eta \rightarrow - \eta \;  \;, \;  \; \theta \rightarrow - \theta \;  \;,
  \;  \; j \rightarrow - j \;  \;, \;  \; \omega \rightarrow \omega^{\star} \;
  \;, \;  \; \beta \rightarrow - \gamma^{\star} \;  \;, \;  \; \gamma
  \rightarrow - \beta^{\star} \; . \label{su3su3conj}
\end{equation}
This explains the apparent discrepancy in the previous section where the
holomorphic three-form carried opposite charge for $\eta = + 1$ F-theory
solutions and their charge conjugate $\eta = - 1$ counterparts, as the
holomorphic three-form is given by $\beta$ (with $\gamma = 0$) in the first
case, and $\gamma^{\star}$ (with $\beta = 0$) in the second, so that charge
conjugation takes $\Psi_-^{\left( 3 \right)} \rightarrow - \Psi_-^{\left( 3
\right)}$ as expected.

\subsection{The covariant ``pure-spinor'' equations} \label{subsec:covarPSE}

We now show how to rewrite the pure spinor equations (\ref{RePsiPl},
\ref{ImPsiPl}, \ref{PsiMn}) as covariant differential conditions on the
chargeless $\tmop{SU}(3) \times \tmop{SU}(3)$
structure. Writing them out rank by rank using (\ref{formstopurespinors}), we
obtain:
\begin{align}
  \mathd \left[ e^{4 A} \eta \right] &= \mathd \alpha - 3 e^{2 A} \tmop{Re}
  \left[ \bar{\mu} \theta \right] \;  \;, \;  \; \mathd \left[ e^{2 A} \theta
  \right] \; = \; 2 i \mu j \;  \;, \;  \; \mathd j \; = \; 0 \;, 
  \label{alphathetajeqn1}\\
  \mathd \left[ e^{2 A + \phi / 2} \tmop{Im} \omega \right] &= \eta e^{4 A}
  H_3 - e^{4 A + \phi} \star \tilde{F}_3 + 3 e^{\phi / 2} \tmop{Im} \left[
  \bar{\mu} (\beta + \bar{\gamma}) \right] \;,  \label{imomegaeqn}\\
  \mathd \left[ i e^{\phi / 2}  \left( \beta + \bar{\gamma} \right) \right] &= e^{2 A} H_3 \wedge \theta + 2 i \mu e^{\phi / 2 - 2 A} \tmop{Im} \omega
  \wedge \tilde{j} \;,  \label{betagammaeqn}\\
  \mathd \left[ e^{\phi}  \left( \frac{1}{2} \eta j^2 + j \wedge j_1 \right)
  \right] &= - e^{2 A + \phi / 2} \tmop{Im} \omega \wedge H_3 - e^{2 \phi}
  \star F_1 - \frac{3}{2} e^{\phi - 2 A} j^2 \wedge \tmop{Re} \left[ \bar{\mu}
  \theta \right] \;,  \label{five1eqn}\\
  \mathd \left[ e^{\phi / 2 - 2 A} \tmop{Im} \omega \wedge \tilde{j} \right] &= j \wedge H_3 \;,  \label{five2eqn}\\
  \mathd \left[ \frac{1}{2} e^{\phi - 2 A} j^2 \wedge \theta \right] &= i
  e^{\phi / 2}  (\beta + \gamma^{\star}) \wedge H_3 + 2 i \mu e^{\phi - 4 A} 
  \frac{1}{6} j^3 \; .  \label{sixformeqn}
\end{align}
The conditions (\ref{alphathetajeqn1}) are already covariant. We decompose the
conditions (\ref{imomegaeqn} -- \ref{sixformeqn}) into covariant pieces,
introducing noncovariant undetermined currents, which we label as ``separation
forms.'' To accomplish this decomposition, we use the following replacements
\begin{eqnarray}
  e^{\phi / 2}  \tilde{F}_3 = \frac{1}{2}  \left( G_3 + G_3^{\star} \right)
  & \;\;,\;\; & e^{- \phi / 2} H_3 = \frac{i}{2}  \left( G_3 - G_3^{\star} \right) \;, \\
  e^{\phi} F_1 = \frac{1}{2 \tau_2}  \left( \mathd \tau
  + \mathd \tau^{\star} \right) & \;\;,\;\; & \mathd \phi = \frac{i}{2
  \tau_2}  \left( \mathd \tau - \mathd \tau^{\star} \right) \;,
\end{eqnarray}
as well as the useful identities
\begin{equation}
\tmop{Im} \left[ \mathrm{D}_+ \xi \right] = e^{- \phi / 2} \mathd \left[
   e^{\phi / 2} \tmop{Im} \xi \right] \;  \;, \;  \; e^{- \phi / 2} \mathd
   \left[ e^{\phi / 2} \xi \right] = \mathrm{D} \xi + \frac{i}{2 \tau_2}
   \mathd \tau \wedge \xi \;,
\end{equation}
for $\xi$ of charge $- 1 / 2$.

The three-form equation (\ref{imomegaeqn}) decomposes into
\begin{equation}
  \mathrm{D}_+ \left[ e^{2 A} \omega \right] = \eta e^{4 A} G_3^{\star} - i
  e^{4 A} \star G_3^{\star} + 3 \bar{\mu} \beta - 3 \mu \gamma +\mathcal{I}_3
  \;, \label{omegasep}
\end{equation}
where $\mathcal{I}_3$ is a real separation form. The four-form equation
(\ref{betagammaeqn}) decomposes into
\begin{align}
  \mathrm{D} \beta - \frac{i}{2 \tau_2} \mathd \tau^{\star} \wedge
  \gamma^{\star} &= - \frac{1}{2} e^{2 A} G_3^{\star} \wedge \theta - i \mu
  e^{- 2 A} \omega \wedge \tilde{j} +\mathcal{J}_4 \;,  \label{betasep}\\
  \mathrm{D} \gamma - \frac{i}{2 \tau_2} \mathd \tau^{\star} \wedge
  \beta^{\star} &= \frac{1}{2} e^{2 A} G_3^{\star} \wedge \bar{\theta} - i
  \bar{\mu} e^{- 2 A} \omega \wedge \tilde{j} -\mathcal{J}_4^{\star} \;, 
  \label{gammasep}
\end{align}
where $\mathcal{J}_4$ is a complex separation form. The first five-form
equation (\ref{five1eqn}) decomposes into:
\begin{align}
  \star \frac{1}{\tau_2} \mathd \tau &= - \frac{i}{\tau_2} \mathd \tau
  \wedge \left( \frac{1}{2} \eta j^2 + j \wedge j_1 \right) + \frac{1}{2} e^{2
  A} \omega^{\star} \wedge G_3 +\mathcal{J}_5 \;,  \label{tausep}\\
  \mathd \left[ \frac{1}{2} \eta j^2 + j \wedge j_1 \right] &= -
  \frac{1}{2} e^{2 A} \tmop{Re} \left[ \omega \wedge G_3 \right] - \frac{3}{2}
  e^{- 2 A} j^2 \wedge \tmop{Re} \left[ \bar{\mu} \theta \right] - \tmop{Re}
  \mathcal{J}_5 \;,  \label{jjsep}
\end{align}
where $\mathcal{J}_5$ is a complex separation form. The second five-form
equation (\ref{five2eqn}) decomposes into:
\begin{equation}
  \mathrm{D}_+  \left[ e^{- 2 A} \omega \wedge \tilde{j} \right] = j \wedge
  G_3^{\star} +\mathcal{I}_5 \;, \label{omegajsep}
\end{equation}
where $\mathcal{I}_5$ is a real separation form. Finally, the six-form
equation (\ref{sixformeqn}) decomposes into:
\begin{align}
  \mathd \left[ \frac{1}{2} e^{- 2 A} j^2 \wedge \theta \right] &=
  \frac{1}{3} i \mu e^{- 4 A} j^3 - \frac{1}{2}  \left( \beta \wedge G_3 -
  \gamma^{\star} \wedge G_3^{\star} \right) + \frac{1}{2}  (\mathcal{J}_6
  -\mathcal{K}_6) \;,  \label{six1sep}\\
  0 &= \gamma^{\star} \wedge G_3 + \frac{i}{2 \tau_2} e^{- 2 A} \mathd \tau
  \wedge j^2 \wedge \theta +\mathcal{J}_6 \;,  \label{six2sep}\\
  0 &= \beta \wedge G_3^{\star} + \frac{i}{2 \tau_2} e^{- 2 A} \mathd
  \tau^{\star} \wedge j^2 \wedge \theta +\mathcal{K}_6 \;,  \label{six3sep}
\end{align}
where $\mathcal{J}_6$ and $\mathcal{K}_6$ are complex separation forms.

To show that the conditions (\ref{omegasep} -- \ref{six3sep}) are covariant,
it is sufficient to prove that all the separation forms $\mathcal{I}_3$,
$\mathcal{J}_4$, $\mathcal{I}_5$, $\mathcal{J}_5$, $\mathcal{J}_6$, and
$\mathcal{K}_6$ must vanish. The derivation is rather technical. We consider
the cases $\eta^2 = 1$ and $\eta^2 < 1$ separately, either one of which must
hold at any point of interest, regardless of whether either is true globally.
In the former case we apply the Hodge and primitivity decompositions with
respect to the local $\tmop{SU}(3)$ structure (\ref{su3extract})
or (\ref{antisu3extract}), and in the latter we decompose with respect to the
local $\tmop{SU}(2)$ structure (\ref{su2extract}). In either
case, applying the $\tmop{SU}(3) \times \tmop{SU}(3)$ stucture constraints (\ref{su3su3cons1} -- \ref{su3su3cons4})
and the covariant conditions (\ref{alphathetajeqn1}), one can show that all
separation forms must vanish. This derivation is summarized in Appendix
\ref{app:derivation}.

We then obtain the explicitly covariant supersymmetry conditions:\footnote{Recall that the forms $\eta$, $\theta$, $j$, $\omega$, $\beta$, and $\gamma$ are equivalent to the chargeless pure spinors by (\ref{purespinorstoforms}, \ref{formstopurespinors}); we prefer them to the pure spinors due to their simpler transformation properties under $\tmop{SL}(2,\mathbb{R})$.}
\begin{align}
  \mathd \left[ e^{4 A} \eta \right] &= \mathd \alpha - 3 e^{2 A} \tmop{Re}
  \left[ \bar{\mu} \theta \right] \;  \;, \;  \; \mathd \left[ e^{2 A} \theta
  \right] =  2 i \mu j \;  \;, \;  \; \mathd j =  0 \;, 
  \label{alphathetajeqn}\\
  \mathrm{D}_+ \left[ e^{2 A} \omega \right] &= \eta e^{4 A} G_3^{\star} -
  i e^{4 A} \star G_3^{\star} + 3 \bar{\mu} \beta - 3 \mu \gamma \;, 
  \label{omegaeqn}\\
  \mathrm{D} \beta - \frac{i}{2 \tau_2} \mathd \tau^{\star} \wedge
  \gamma^{\star} &= - \frac{1}{2} e^{2 A} G_3^{\star} \wedge \theta - i \mu
  e^{- 2 A} \omega \wedge \tilde{j} \;,  \label{betaeqn}\\
  \mathrm{D} \gamma - \frac{i}{2 \tau_2} \mathd \tau^{\star} \wedge
  \beta^{\star} &= \frac{1}{2} e^{2 A} G_3^{\star} \wedge \bar{\theta} - i
  \bar{\mu} e^{- 2 A} \omega \wedge \tilde{j} \;,  \label{gammaeqn}\\
  \mathrm{D}_+  \left[ e^{- 2 A} \omega \wedge \tilde{j} \right] &= j
  \wedge G_3^{\star} \;,  \label{omegajeqn}\\
  \mathd \left[ \frac{1}{2} e^{- 2 A} j^2 \wedge \theta \right] &=
  \frac{1}{3} i \mu e^{- 4 A} j^3 - \frac{1}{2}  \left( \beta \wedge G_3 -
  \gamma^{\star} \wedge G_3^{\star} \right) \;,  \label{six1eqn}
\end{align}
as well as
\begin{align}
  \star \frac{1}{\tau_2} \mathd \tau &= - \frac{i}{\tau_2} \mathd \tau
  \wedge \left( \frac{1}{2} \eta j^2 + j \wedge j_1 \right) + \frac{1}{2} e^{2
  A} \omega^{\star} \wedge G_3 \;,  \label{taueqn}\\
  \mathd \left[ \frac{1}{2} \eta j^2 + j \wedge j_1 \right] &= -
  \frac{1}{2} e^{2 A} \tmop{Re} \left[ \omega \wedge G_3 \right] - \frac{3}{2}
  e^{- 2 A} j^2 \wedge \tmop{Re} \left[ \bar{\mu} \theta \right] \;, 
  \label{jjeqn}\\
  0 &= \gamma^{\star} \wedge G_3 + \frac{i}{2 \tau_2} e^{- 2 A} \mathd \tau
  \wedge j^2 \wedge \theta \;,  \label{six2eqn}\\
  0 &= \beta \wedge G_3^{\star} + \frac{i}{2 \tau_2} e^{- 2 A} \mathd
  \tau^{\star} \wedge j^2 \wedge \theta \; .  \label{six3eqn}
\end{align}
In fact, (\ref{alphathetajeqn} -- \ref{six1eqn}), together with the algebraic constraints (\ref{su3su3cons1} -- \ref{su3su3cons4}), imply the remaining equations, (\ref{taueqn} -- \ref{six3eqn}). This redundancy is not surprising, given that five noncovariant equations (\ref{imomegaeqn} -- \ref{sixformeqn})
have been shown to imply, taken together with (\ref{su3su3cons1} -- \ref{su3su3cons4}, \ref{alphathetajeqn1}), nine covariant equations (\ref{omegaeqn} -- \ref{six3eqn}).

To establish this last result, we follow similar steps to those taken to derive the covariant conditions. We add arbitrary forms $\mathcal{J}_5$, $\mathcal{M}_5$, $\mathcal{J}_6$, and $\mathcal{K}_6$ respectively to (\ref{taueqn} -- \ref{six3eqn}) and then show that these forms must vanish upon imposing the other supersymmetry conditions. The math is now very similar
to that used to derive the covariant conditions. In particular, for $\eta^2 < 1$, the same steps
that led to (\ref{ibetagammaeqn1}, \ref{ibetagammaeqn2}) show that
$\mathcal{J}_6 =\mathcal{K}_6 = 0$ as a consequence of (\ref{omegaeqn},
\ref{betaeqn}, \ref{gammaeqn}). Similarly, the steps which led to
(\ref{J5eqn}) and (\ref{cond5}, \ref{cond7}) show that $\mathcal{J}_5 = 0$,
and those which led to (\ref{ibetagammaeqn3}) and (\ref{cond6}) show that
$\mathcal{M}_5 = 0$. The special case $\eta = \pm 1$ is readily verified.

Even so, the redundant conditions (\ref{taueqn} -- \ref{six3eqn}) are sometimes useful in computations.

\section{Consistency Checks} \label{sec:consistency}

Having established our main results, the distinction between charged and chargeless solutions, the $\tmop{SL}(2,\mathbb{R})$ transformation properties of the chargeless pure spinors (\ref{formstopurespinors}), and the $\tmop{SL}(2,\mathbb{R})$-covariant supersymmetry conditions (\ref{alphathetajeqn} -- \ref{six1eqn}), we now perform a few additional computations as consistency checks on these results. In \S\ref{subsec:eoms}, we review the supergravity Bianchi identities, and show that they imply the flux equations of motion upon imposition of the supersymmetry conditions, a known result which we are able to rederive relatively easily. In \S\ref{subsec:superpotential}, we show that the flux superpotential proposed in~\cite{Koerber:2007xk} is $\tmop{SL}(2,\mathbb{R})$ invariant, a new result which presents a further consistency check on our calculation and on the proposed superpotential.

\subsection{Equations of motion} \label{subsec:eoms}

In addition to the supersymmetry conditions, four-dimensional $\mathcal{N}= 1$
vacua must satisfy the supergravity Bianchi identities\footnote{Our distinction between ``Bianchi identities'' and ``equations of motion'' is consistent with that made in~\cite{Grana:2005sn}; the RR Bianchi identity $\mathd F_1 = 0$ is implicit in this formalism, since $\tau$ is specified directly.}
\begin{equation}
\mathd \left[ e^{- 8 A} \star \mathd \alpha \right] = - \frac{i}{2} G_3 \wedge G_3^{\star} \;\;,\;\;\mathrm{D}_- G_3 = 0 \;, \label{sugraBianchis}
\end{equation}
and equations of motion:
\begin{align}
  \mathd \star \mathd A &= \frac{1}{8} e^{4 A} G_3
  \wedge \star G_3^{\star} + \frac{1}{4} e^{- 8 A} \mathd \alpha \wedge
  \star \mathd \alpha + \Lambda e^{- 4 A} \Omega_6 \;,  \label{sugraeqns1} \\
  \mathrm{D}_+ \left[ e^{4 A} \star G_3 \right] &= i \mathd \alpha \wedge G_3
  \;\;,\;\; \mathrm{D} \left[ \star \frac{1}{\tau_2} \mathd \tau \right] = -
  \frac{i}{2} e^{4 A} G_3 \wedge \star G_3 \;, \label{sugraeqns2}
\end{align}
\begin{equation}
  R_{m n} = 8 \nabla_m A \nabla_n A - \frac{1}{2} e^{- 8 A} \nabla_m
  \alpha \nabla_n \alpha + \frac{1}{4 \tau_2^2}  \left[ \nabla_m \tau \nabla_n
  \bar{\tau} + c.c. \right] + \frac{1}{2} e^{4 A}  \hat{T}_{m n} - \Lambda
  e^{- 4 A} g_{m n} \;, \label{Einsteqn}
\end{equation}
where $\Lambda =\mathcal{R}_{(4)} / 4$ is the four-dimensional cosmological
constant, $R_{m n}$ is the Ricci tensor formed from the unwarped metric $g_{m
n}$, contractions and Hodge duals are formed using $g_{m n}$, and
\begin{equation}
  \hat{T}^m_n = \frac{1}{4}  \left( G^{mpq}  \bar{G}_{npq} + \bar{G}^{mpq}
  G_{npq} \right) - \frac{1}{12}  \bar{G}^{pqr} G_{pqr} \delta^m_n \; .
\end{equation}

Fortunately, one can show that the supersymmetry conditions,
combined with the Bianchi identities (\ref{sugraBianchis}), imply the remaining equations of motion (\ref{sugraeqns1}, \ref{sugraeqns2}, \ref{Einsteqn})~\cite{Gauntlett:2005ww}.
This result can be extended to include calibrated D-branes, wherein the bulk
supersymmetry conditions and Bianchi identities, together with the calibration
equations, imply all remaining bulk and brane equations of motion~\cite{Koerber:2007hd}.

For completeness, we partially reproduce this result for the chargeless solutions
considered here. While we do not impose calibration conditions, we do not
exclude sources explicitly, and do not impose the source-free Bianchi
identities in the following derivation, apart from the Bianchi identity
$\mathrm{D}  \left[ \frac{1}{\tau_2} \mathd \tau \right] =
0$.{\footnote{Violating the Bianchi identity $\mathrm{D}  \left[
\frac{1}{\tau_2} \mathd \tau \right] = 0$ in an $\tmop{SL} (2, \mathbb{Z})$
covariant formalism requires $\tmop{SL} (2, \mathbb{R})$ to be gauged and spontaneously broken to $\tmop{SL} (2, \mathbb{Z})$,
which is beyond the scope of this paper. This gauging is necessary in the
vicinity of a seven-brane, due to the topological defect caused by the
monodromy.}}

Applying $\mathrm{D}_+$ to (\ref{omegaeqn}) and simplifying using
(\ref{betaeqn}, \ref{gammaeqn}, \ref{alphathetajeqn}), we find:
\begin{equation}
  \mathrm{D}_+ \left[ e^{4 A} \star G_3 \right] = i \mathd \alpha \wedge G_3 +
  i \eta e^{4 A}  \mathrm{D}_- G_3
\end{equation}
Thus, in the absence of sources, the $G_3$ Bianchi identity implies the $G_3$
equation of motion. Applying $\mathrm{D}$ to (\ref{taueqn}) and simplifying
using (\ref{jjeqn}, \ref{omegaeqn}, \ref{six2eqn}, \ref{six3eqn}), we find:
\begin{equation}
  \mathrm{D} \left[ \star \frac{1}{\tau_2} \mathd \tau \right] = - \frac{i}{2}
  e^{4 A} G_3 \wedge \star G_3 + \frac{1}{2} e^{2 A} \omega^{\star} \wedge
  \mathrm{D}_- G_3
\end{equation}
Thus, in the source-free case, the axodilaton equation of motion also follows
from the $G_3$ Bianchi identity.

To obtain the warp-factor equation of motion from the Bianchi identities, we
use the identity
\begin{equation}
  0 = \frac{1}{2} e^{- 6 A} \star \mathd \left[ (1 - \eta^2) e^{8 A} \right] -
  \frac{1}{2} e^{4 A} \tmop{Im} \left[ \omega \wedge G_3 \right] - \frac{3}{2}
  \eta j^2 \wedge \tmop{Im} \left[ \bar{\mu} \theta \right]
\end{equation}
which can be shown to follow from the supersymmetry conditions. Combining this
with $\eta e^{- 4 A}$ times the Hodge star of (\ref{alphathetajeqn}), we obtain:
\begin{equation}
  0 = e^{- 4 A} \star \mathd \left[ e^{4 A} \right] - \frac{1}{2} e^{2 A}
  \tmop{Im} \left[ \omega \wedge G_3 \right] - \eta e^{- 4 A} \star \mathd
  \alpha
\end{equation}
Taking the exterior derivative of this equation, and applying (\ref{omegaeqn},
\ref{jjeqn}, \ref{six1eqn}), we obtain:
\begin{align}
  0 &= 4 \mathd \star \mathd A - \frac{1}{2} e^{4 A} G_3 \wedge \star
  G_3^{\star} - e^{- 8 A} \mathd \alpha \wedge \star \mathd \alpha + 12 | \mu
  |^2 e^{- 4 A}  \left[ \frac{1}{6} \eta j^3 + \frac{1}{2} j^2 \wedge j_1
  \right] \nonumber\\
  &\hspace{1cm} - \eta e^{4 A}  \left[ \mathd \left[ e^{- 8 A} \star \mathd \alpha
  \right] + \frac{i}{2} G_3 \wedge G_3^{\star} \right] - \frac{1}{2} e^{2 A}
  \tmop{Im} \left[ \omega \wedge \mathrm{D}_- G_3 \right] 
\end{align}
The first line is the source-free $A$ equation of motion with cosmological
constant $\Lambda = - 3 | \mu |^2$. Thus, this too follows from the Bianchi
identities in the absence of sources.

The chargeless supersymmetry conditions also impose constraints upon
$\mathrm{D}_- G_3$ itself. Applying $\mathrm{D}$ to (\ref{betaeqn},
\ref{gammaeqn}) and simplifying using (\ref{omegajeqn}), and applying
$\mathrm{D}_+$ to (\ref{omegajeqn}) itself, we find:
\begin{equation}
  \mathrm{D}_- G_3 \wedge \theta = \mathrm{D}_- G_3 \wedge \bar{\theta} =
  \mathrm{D}_- G_3 \wedge j = 0
\end{equation}
Moreover, taking the exterior derivative of (\ref{jjeqn}) and simplifying
using (\ref{omegaeqn}, \ref{six1eqn}, \ref{taueqn}), we obtain:
\begin{equation}
  \tmop{Re} \left[ \omega \wedge \mathrm{D}_- G_3 \right] = 0
\end{equation}
These equations constrain the form of possible source terms consistent with
$\mathcal{N}= 1$ supersymmetry, though such questions are more thoroughly
addressed by the study of D-brane calibrations~\cite{Koerber:2005qi,Martucci:2005ht,Martucci:2006ij}.

While it is possible to derive the Einstein equations (\ref{Einsteqn}) from
the supersymmetry conditions and Bianchi identities~\cite{Gauntlett:2005ww}, we do not
attempt to reproduce such a computation in this context.

\subsection{The flux superpotential}\label{subsec:superpotential}

The off-shell flux superpotential
\footnote{Specifically, this is the superpotential of four-dimensional ``Weyl-invariant supergravity,'' as explained in detail in~\cite{Koerber:2007xk}. The more-standard Einstein supergravity superpotential is then $W_E = e^{-K/2}\, W$, where $K$ is the K\"{a}hler potential. I thank P.~C\'{a}mara for discussions on this point.}
\begin{equation}
W=\frac{1}{4 \kappa_{10}^2} \int \left< e^{3 A^{(S)}-\phi} \Psi_-^{(S)},  F+i\, \mathd_H \left[e^{-\phi} \tmop{Re} \Psi_+^{(S)} \right] \right> \; , \label{KMsuperpotential}
\end{equation}
has been proposed~\cite{Benmachiche:2006df, Koerber:2007xk} as
an appropriate generalization of the well-known Gukov-Vafa-Witten superpotential~\cite{Gukov:1999ya, Taylor:1999ii} to general $\tmop{SU}(3)\times\tmop{SU(3)}$ structure compactifications with equal spinor norms ($k_1=0$). Since $\tmop{SL}_{\pm}(2,\mathbb{Z})$ is an exact gauged symmetry of string theory, $W$ must be $\tmop{SL}_{\pm}(2,\mathbb{Z})$ invariant. As usual, we expect that this is enhanced to $\tmop{SL}_{\pm}(2,\mathbb{R})$ invariance in tree-level supergravity, so that the integrand of (\ref{KMsuperpotential}) must be neutral under $\tmop{SL}(2,\mathbb{R})$.

We now verify that this is the case for chargeless solutions. Applying the chargeless ansatz (\ref{formstopurespinors}) to (\ref{KMsuperpotential}) and simplifying the integrand using (\ref{su3su3cons1} -- \ref{su3su3cons4}), we obtain:
\begin{align}
W &= \frac{i}{4 \kappa_{10}^2} \int \left(G_3\wedge\beta+G_3^{\star}\wedge\gamma^{\star}
             +\frac{1}{2} \beta\wedge\mathrm{D} [e^{-2 A} \omega^{\star} ]
              -\frac{1}{2} \gamma^{\star} \wedge\mathrm{D} [e^{-2 A} \omega ]\right) \nonumber \\
&\hspace{1cm}
            +\frac{1}{4 \kappa_{10}^2} \int \left(e^{2 A}\, \theta\wedge\tilde{F}_5
             -i\, \theta\wedge j \wedge \mathd[\eta e^{-2 A} j] - i\, e^{-2 A} j \wedge j_1\wedge \mathd\theta \right) \;. \label{covarsuperpotential}
\end{align}
Comparing with (\ref{su3su3conj}), we see that the superpotential is $\tmop{SL}_{\pm}(2,\mathbb{R})$ invariant. Moreover, for $\eta=1$, it truncates to the familiar Gukov-Vafa-Witten result:
\begin{equation}
W = \frac{i}{4 \kappa_{10}^2} \int G_3\wedge\Omega \; ,
\end{equation}
where $\Omega = \beta$ is the holomorphic three-form.

Applying the supersymmetry conditions (\ref{alphathetajeqn} -- \ref{six3eqn}) to (\ref{covarsuperpotential}) and simplifying the integrand, we find
\begin{equation}
\left< W \right > = \frac{\mu}{\kappa_{10}^2} \int \mathd^6 y \sqrt{g} e^{-4 A} = \frac{\mu}{\kappa_4^2} \;, \label{superpotentialvev}
\end{equation}
for a supersymmetric vacuum, where $\kappa_4^2$ is the four-dimensional Newton constant. This is consistent with the supergravity result $\Lambda = -3 \kappa_4^4 |W|^2 = -3 |\mu|^2$.

The $\tmop{SL}(2,\mathbb{R})$ covariance of (\ref{covarsuperpotential}) is a highly non-trivial consistency check on the proposed superpotential~(\ref{KMsuperpotential}), as the latter was developed using D-brane and Euclidean D-brane physics~\cite{Koerber:2007xk} without imposing $\tmop{SL}(2,\mathbb{Z})$ invariance.

\section{Conclusions}\label{sec:conclusions}

We have shown that geometric $\mathcal{N}= 1$ vacua of type IIB string theory
fall into two classes, which we label chargeless and charged solutions.
Chargeless solutions are particularly interesting from the perspective of
F-theory, as they allow in-principle arbitrary combinations of $\tmop{SL}(2, \mathbb{Z})$ monodromies. We have derived simple algebraic
(\ref{su3su3cons1} -- \ref{su3su3cons4}) and differential (\ref{alphathetajeqn} --
\ref{six1eqn}) conditions for chargeless supersymmetric solutions which are
manifestly $\tmop{SL} (2, \mathbb{R})$ covariant. Together with the Bianchi identities (\ref{sugraBianchis}), these are necessary and sufficient conditions for chargeless supersymmetry. The success of
this endeavor is a non-trivial consistency check on the pure-spinor equations
of~\cite{Grana:2005sn}, which do not make the $\tmop{SL} (2, \mathbb{R})$
invariance of the theory manifest.

We have also demonstrated that the flux superpotential proposed in~\cite{Koerber:2007xk} is $\tmop{SL}_{\pm}(2,\mathbb{R})$ invariant for chargeless $\tmop{SU}(3)\times\tmop{SU}(3)$ structure, obtaining the covariant expression (\ref{covarsuperpotential}).

The formalism presented here should prove useful to the study of generalized F-theory solutions, where $\tmop{SL}(2,\mathbb{Z})$ covariance plays an essential role. It also provides a useful alternative perspective on previous approaches to the classification of $\mathcal{N}=1$ vacua using generalized complex geometry.

One might hope to extend these methods to charged solutions. Indeed, in the
case of strict $\tmop{SU}(3)$ structure, the calculation is
relatively straightforward, and results in a clean restatement of the
supersymmetry conditions on an already well-studied class of vacua. There are
indications that the $\tmop{SL} (2, \mathbb{R})$-covariant supersymmetry
conditions on general charged vacua should be relatively simple, but an
explicit derivation of these conditions is hampered by the difficulty in
determining the $\tmop{SL} (2, \mathbb{R})$ transformation
properties of the pure spinors, since the considerations of
{\S}\ref{subsec:chargelessSU3xSU3structure} no longer apply. A more direct approach using the known $\tmop{SL}(2,\mathbb{R})$ transformation law for the supersymmetry generators $\epsilon^i$ may be indicated. We return to these questions in a future work~\cite{future}.

\section*{Acknowledgements}

I would like to thank M.~Berg, P.~C\'{a}mara, P.~Koerber, L.~McAllister, G.~Torroba, and T.~Wrase for
useful discussions and A.~Dymarsky, P.~Koerber, L.~McAllister, A.~Tomasiello, G.~Torroba, and T.~Wrase for comments on the
manuscript. My research was funded in part by the NSF under grant PHY-0757868.
I am grateful to the Swedish Foundation for International Cooperation in
Research and Higher Education for their support while part of this research
was carried out.

\appendix\section{Derivation of the chargeless SUSY
conditions}\label{app:derivation}

In \S\ref{subsec:covarPSE}, we showed that the pure spinor equations (\ref{RePsiPl}, \ref{ImPsiPl}, \ref{PsiMn}) can be rewritten in the form (\ref{alphathetajeqn1}, \ref{omegasep} -- \ref{six3sep}) for arbitrary real separation forms $\mathcal{I}_3$ and $\mathcal{I}_5$ and complex separation forms $\mathcal{J}_4$, $\mathcal{J}_5$, $\mathcal{J}_6$ and $\mathcal{K}_6$. We now show that these separation forms all vanish, proving that the supersymmetry conditions are covariant.

We consider the cases $\eta^2 = 1$ and $\eta^2 < 1$ separately, in
{\S}\ref{subsec:su3loci} and {\S}\ref{subsec:localsu2} respectively.

\subsection{$\tmop{SU}(3)$ structure loci $\left( \eta = \pm 1
\right)$}\label{subsec:su3loci}

We first prove that the separation forms vanish at a locus where $\eta^2 = 1$.
We consider the case $\eta = + 1$ ($\eta = - 1$ is related to this by charge
conjugation). Note that the $\tmop{SU}(3) \times \tmop{SU}(3)$ structure
constraints (\ref{su3su3cons1} -- \ref{su3su3cons4}) and
(\ref{alphathetajeqn1}) imply the conditions:
\begin{eqnarray}
  \mathd \theta = 2 i \mu e^{- 2 A} j \;  \;, & \mathd \eta = 0 & , \;  \;
  \mathd \gamma = 0 \;, \\
  \left[ \mathd \omega \right]_{(0, 3)} = \left[ \mathd \omega \right]_{(1,
  2)} = 0 \;  \;, & j \wedge \mathd \omega = 0 & , \;  \; \mathd \omega \wedge
  \beta^{\star} = - 4 i \bar{\mu} e^{- 2 A} j^3 \;, 
\end{eqnarray}
where $\mathrm{D} \omega = \mathrm{D}_+ \omega = \mathd \omega$, since $\omega
= 0$, so that the connection terms vanish, and the Hodge decomposition is
taken with respect to the locally defined almost complex structure $\beta$.
This complex structure need not be integrable. However, the $(1, 2)$ component
of $\mathd \omega$ must still vanish, since $\omega$ is a $(2, 0)$ form which
vanishes where $\eta = 1$, so that $\omega \wedge f_{\left( 2, 1 \right)} = 0$
for any $(2, 1)$ form $f$. Taking the exterior derivative and imposing $\omega
= 0$, we recover $\left[ \mathd \omega \right]_{(1, 2)} = 0$ since $f$ is
arbitrary.

Written out, (\ref{omegasep} -- \ref{six3sep}) reduce to:
\begin{eqnarray}
  e^{2 A} \mathd \omega = e^{4 A}  [G_3^{\star} - i \star G_3^{\star}] + 3
  \bar{\mu} \beta +\mathcal{I}_3 \;  \;, & \mathrm{D} \beta =\mathcal{J}_4 & ,
  \;  \; \frac{i}{2 \tau_2} \mathd \tau^{\star} \wedge \beta^{\star}
  =\mathcal{J}_4^{\star} \;,  \label{su3sep1}\\
  \star \frac{1}{\tau_2} \mathd \tau = - \frac{i}{2 \tau_2} \mathd \tau \wedge
  j^2 +\mathcal{J}_5 \;  \;, & \tmop{Re} \mathcal{J}_5 = 0 & , \;  \; j \wedge
  G_3^{\star} +\mathcal{I}_5 = 0 \;,  \label{su3sep2}\\
  0 = \frac{2}{3} i \mu e^{- 4 A} j^3 + \frac{1}{2} \beta \wedge G_3 -
  \frac{1}{2}  (\mathcal{J}_6 -\mathcal{K}_6) \;  \;, & \mathcal{J}_6 = 0 & ,
  \;  \; \beta \wedge G_3^{\star} +\mathcal{K}_6 \; .  \label{su3sep3}
\end{eqnarray}
Wedging $\beta^{\star}$ into the first equation of (\ref{su3sep1}) and
simplifying, we find $\mathcal{I}_3 \wedge \beta^{\star} = 0$. Therefore,
since $\mathcal{I}_3$ is real, $\mathcal{I}_3 = (2, 1) \oplus (1, 2)$.
However, the rest of equation only has $(3, 0) \oplus (2, 1)^{\tmop{P}} \oplus (1,
2)^{\tmop{NP}} \oplus (0, 3)$ components.\footnote{The superscripts $\tmop{P}$ and $\tmop{NP}$ denote primitive and non-primitive components.} Since $\mathcal{I}_3$ is real, this
implies that it must vanish, and therefore in particular $G_{(2,
1)}^{\tmop{NP}} = G_{(3, 0)} = 0$. Thus, $\mathcal{K}_6 = 0$ and
$\mathcal{I}_5 = j \wedge (2, 1)^{\tmop{NP}}$ and is therefore vanishing,
since it is real.

Writing out the Hodge star in the first equation of (\ref{su3sep2}), we find:
\begin{equation}
  \frac{i}{2 \tau_2} j^2 \wedge \bar{\partial} \tau = \mathcal{J}_5 
\end{equation}
However, since $\mathcal{J}_5$ is imaginary, we conclude that it must vanish,
and therefore $\bar{\partial} \tau = 0$. Applying this to (\ref{su3sep1}), we
find $\mathcal{J}_4 = 0$. Thus, all the separation forms must vanish at a
locus where $\eta = 1$. A similar argument applies to the case $\eta = - 1$.

\subsection{Local $\tmop{SU}(2)$ structure ($\eta^2 <
1$)}\label{subsec:localsu2}

Now consider a point where $\eta^2 < 1$; we can define a local $\tmop{SU}(2)$
structure $J_2$, $\Omega_2$, and $\Theta$ via (\ref{su2extract}). Using this
$\tmop{SU}(2)$ structure, we can decompose an arbitrary forms according to
their $\theta$ and $\bar{\theta}$ fiber components, as well as their Hodge
type and (for $(1, 1)$ forms) their primitivity along the base. Thus, for
instance, an arbitrary three-form decomposes as
\begin{align}
  M_3 &= M_{\theta \bar{\theta} ; 1, 0} + M_{\theta \bar{\theta} ; 0, 1} +
  M_{; 2, 1} + M_{; 1, 2} + M_{\theta ; 2, 0} + M_{\theta ; 0, 2} +
  M_{\bar{\theta} ; 2, 0} + M_{\bar{\theta} ; 0, 2}  \nonumber\\
  &\hspace{1cm} + M_{\theta ; (1, 1)^{\tmop{P}}} +M_{\bar{\theta} ; (1, 1)^{\tmop{P}}}+ M_{\theta ; (1, 1)^{\tmop{NP}}} + M_{\bar{\theta} ; (1,1)^{\tmop{NP}}} \;, 
\end{align}
and an arbitrary four-form as
\begin{equation}
  N_4 = N_{; 2, 2} + N_{\theta \bar{\theta} ; (1, 1)^{\tmop{NP}}} + N_{\theta
  \bar{\theta} ; (1, 1)^{\tmop{P}}} + N_{\theta \bar{\theta} ; 2, 0} + N_{\theta
  \bar{\theta} ; 0, 2} + N_{\theta ; 2, 1} + N_{\theta ; 1, 2} +
  N_{\bar{\theta} ; 2, 1} + N_{\bar{\theta} ; 1, 2} \; .
\end{equation}
Note the use of the semicolon to distinguish this from an ordinary Hodge
decomposition; e.g. $M_{; 1, 2} \neq M_{(1, 2)}$, since the former has legs
along the base only. Many of these components can be written as scalars times
the $\tmop{SU}(2)$ structure forms, for instance $M_{\theta ; (1,
1)^{\tmop{NP}}} \propto \Theta \wedge J_2$ and $N_{; 2, 2} \propto J_2^2$,
etc.

We use these decompositions to show that the separation forms vanish. To begin
with, we consider the $\theta \wedge (1, 1)^{\tmop{P}}$ and $\bar{\theta} \wedge (1,
1)^{\tmop{P}}$ components of (\ref{omegasep}), along with the $J_1 \wedge (1, 1)^{\tmop{P}}$
components of (\ref{betasep}, \ref{gammasep}):
\begin{align}
  e^{2 A}  \left[ \mathrm{D} \omega \right]_{\theta ; (1, 1)^{\tmop{P}}} &= (1 +
  \eta) e^{4 A}  \left[ G_{\bar{\theta} ; (1, 1)^{\tmop{P}}} \right]^{\star}
  +\mathcal{I}_{\theta ; (1, 1)^{\tmop{P}}} \;,  \label{thetaprim1}\\
  e^{2 A}  \left[ \mathrm{D} \omega \right]_{\bar{\theta} ; (1, 1)^{\tmop{P}}} &= -
  (1 - \eta) e^{4 A}  \left[ G_{\theta ; (1, 1)^{\tmop{P}}} \right]^{\star}
  +\mathcal{I}_{\bar{\theta} ; (1, 1)^{\tmop{P}}} \;,  \label{thetaprim2}\\
  \left[ \mathrm{D} \omega \right]_{\bar{\theta} ; (1, 1)^{\tmop{P}}} \wedge \theta &=
  - e^{2 A}  (1 - \eta)  \left[ G_{\theta ; (1, 1)^{\tmop{P}}} \right]^{\star} \wedge
  \theta +  2 (1 - \eta) \mathcal{J}_{\theta \bar{\theta} ; (1,
  1)^{\tmop{P}}} \;,  \label{thetaprim3}\\
  \left[ \mathrm{D} \omega \right]_{\theta ; (1, 1)^{\tmop{P}}} \wedge \bar{\theta} &=
  e^{2 A}  (1 + \eta)  \left[ G_{\bar{\theta} ; (1, 1)^{\tmop{P}}} \right]^{\star}
  \wedge \bar{\theta} - 2 (1 + \eta)  \left[ \mathcal{J}_{\theta \bar{\theta}
  ; (1, 1)^{\tmop{P}}} \right]^{\star} \; .  \label{thetaprim4}
\end{align}
Wedging $\bar{\theta}$ and $\theta$ into (\ref{thetaprim1}) and
(\ref{thetaprim2}) respectively, and combining them with (\ref{thetaprim4})
and (\ref{thetaprim3}) to eliminate $G_3$, we find
\begin{equation}
  0 = e^{- 2 A} \mathcal{I}_{\theta ; (1, 1)^{\tmop{P}}} \wedge \bar{\theta} + 2 (1 +
  \eta)  \left[ \mathcal{J}_{\theta \bar{\theta} ; (1, 1)^{\tmop{P}}} \right]^{\star}
  \;  \;, \;  \; 0 = e^{- 2 A} \mathcal{I}_{\bar{\theta} ; (1, 1)^{\tmop{P}}} \wedge
  \theta -  2 (1 - \eta) \mathcal{J}_{\theta \bar{\theta} ; (1,
  1)^{\tmop{P}}} \; . \label{IJprimeqn}
\end{equation}
Using the reality of $\mathcal{I}_3$, we deduce that $\mathcal{I}_{\theta ;
\left( 1, 1 \right)^{\tmop{P}}} =\mathcal{I}_{\bar{\theta} ; \left( 1, 1 \right)^{\tmop{P}}}
=\mathcal{J}_{\theta \bar{\theta} ; \left( 1, 1 \right)^{\tmop{P}}} = 0$.

We extract further components of the separation forms by wedging them into
various of the $\tmop{SU}(2)$ structure forms. Wedging $j \wedge \theta$ and
$j \wedge \bar{\theta}$ into (\ref{omegasep}) and combining with $\theta$ and
$\bar{\theta}$ wedged into (\ref{omegajsep}), we find that $\mathcal{I}_5
\wedge \theta =\mathcal{I}_3 \wedge j \wedge \theta = 0$, as well as $G_3
\wedge j \wedge \theta = G_3 \wedge j \wedge \bar{\theta} = 0$. Wedging $j$
and $\theta$ into (\ref{betasep}), we obtain $\mathcal{J}_4 \wedge \theta
=\mathcal{J}_4 \wedge j = 0$.

Now consider $\beta$ and $\gamma$ wedged into (\ref{omegasep}). Integrating by
parts, applying (\ref{betasep}, \ref{gammasep}), and using (\ref{six2sep},
\ref{six3sep}) to eliminate $G_3$, we obtain:
\begin{equation}
  \mathcal{I}_3 \wedge \beta + e^{2 A} \mathcal{J}_4 \wedge \omega + 2 \eta
  e^{4 A} \mathcal{K}_6 = 0 \;  \;, \;  \; \mathcal{I}_3 \wedge \gamma - e^{2
  A} \mathcal{J}_4^{\star} \wedge \omega + 2 \eta e^{4 A}
  \mathcal{J}_6^{\star} = 0 \;, \label{ibetagammaeqn1}
\end{equation}
where we make use of the identities
\begin{equation}
  \omega^{\star} \wedge \beta = (1 + \eta) j^2 \wedge \theta \;  \;, \;  \;
  \omega \wedge \gamma^{\star} = (1 - \eta) j^2 \wedge \theta \;,
\end{equation}
and
\begin{equation}
  \star \theta = - \frac{i}{2} j^2 \wedge \theta \;  \;, \;  \; \star \beta =
  - i \beta \;  \;, \;  \; \star \gamma = i \gamma \;  \;, \;  \; \star (j
  \wedge \theta) = - i j \wedge \theta
\end{equation}
We also consider $\frac{1}{2} \omega \wedge \theta$ and $\frac{1}{2} \omega
\wedge \bar{\theta}$ wedged into (\ref{omegasep}). Integrating by parts and
using (\ref{six2sep}, \ref{six3sep}) to eliminate $G_3$, we obtain:
\begin{equation}
  \mathcal{I}_3 \wedge \beta + (1 + \eta) e^{4 A} \mathcal{K}_6 = 0 \;  \;, \;
  \; \mathcal{I}_3 \wedge \gamma - (1 - \eta) e^{4 A} \mathcal{J}_6^{\star} =
  0 \label{ibetagammaeqn2}
\end{equation}
where we cancel an overall factor of $(1 - \eta)$ from the first equation and
$(1 + \eta)$ from the second; these equations still hold in the special case
$\eta = \pm 1$, since they then follow from (\ref{ibetagammaeqn1}).

Wedging $\theta$ and $\bar{\theta}$ into (\ref{tausep}) and using
(\ref{six2sep}, \ref{six3sep}) to eliminate $G_3$ as before, we obtain:
\begin{equation}
  \mathcal{J}_5 \wedge \theta + (1 + \eta) e^{2 A} \mathcal{J}_6 = 0 \;  \;,
  \;  \; \mathcal{J}_5 \wedge \bar{\theta} + (1 - \eta) e^{2 A}
  \mathcal{K}_6^{\star} = 0 \; . \label{J5eqn}
\end{equation}
Next, consider (\ref{jjsep}) wedged into $\theta$:
\begin{equation}
  0 = - i \mu j^2 \wedge j_1 + e^{2 A} \mathd \eta \wedge j^2 \wedge \theta^{}
  + e^{4 A}  (1 - \eta) G_3 \wedge \beta + e^{4 A}  (1 + \eta) G_3^{\star}
  \wedge \gamma^{\star} + 2 e^{2 A} \tmop{Re} \mathcal{J}_5 \wedge \theta \; .
  \label{sixformalt}
\end{equation}
We compare this with the wedge product of $\beta^{\star}$ and $\gamma^{\star}$
with (\ref{omegasep}). Integrating by parts, applying (\ref{betasep},
\ref{gammasep}), and simplifying, we obtain:
\begin{equation}
  - i \bar{\mu} j^2 \wedge j_1 - e^{2 A} \mathd \eta \wedge j^2 \wedge
  \bar{\theta} - (1 + \eta) e^{4 A} G_3 \wedge \gamma - (1 - \eta) e^{4 A}
  G_3^{\star} \wedge \beta^{\star} + e^{2 A} \omega \wedge
  \mathcal{J}_4^{\star} +\mathcal{I}_3 \wedge \beta^{\star} = 0 \;,
\end{equation}
\begin{equation}
  - i \mu j^2 \wedge j_1 + e^{2 A} \mathd \eta \wedge j^2 \wedge \theta + (1 -
  \eta) e^{4 A} G_3 \wedge \beta + (1 + \eta) e^{4 A} G_3^{\star} \wedge
  \gamma^{\star} - e^{2 A} \omega \wedge \mathcal{J}_4 +\mathcal{I}_3 \wedge
  \gamma^{\star} = 0 \;,
\end{equation}
where we use
\begin{equation}
  2 (1 + \eta) j^3 + \omega \wedge \omega^{\star} \wedge \tilde{j} - 3 i \beta
  \wedge \beta^{\star} = - 2 (1 - \eta) j^3 + \omega \wedge \omega^{\star}
  \wedge \tilde{j} + 3 i \gamma \wedge \gamma^{\star} = - j^2 \wedge j_1 \;,
\end{equation}
which can be verified a number of different ways. Thus, we find:
\begin{equation}
  \mathcal{I}_3 \wedge \beta^{\star} + e^{2 A} \omega \wedge
  \mathcal{J}_4^{\star} + 2 e^{2 A} \tmop{Re} \mathcal{J}_5 \wedge
  \bar{\theta} = 0 \;  \;, \;  \; \mathcal{I}_3 \wedge \gamma^{\star} - e^{2
  A} \omega \wedge \mathcal{J}_4 - 2 e^{2 A} \tmop{Re} \mathcal{J}_5 \wedge
  \theta = 0 \; . \label{ibetagammaeqn3}
\end{equation}
Finally, consider $\frac{1}{2} \omega^{\star} \wedge \theta$ wedged into
(\ref{omegasep}). Integrating by parts and simplifying, we obtain:
\begin{align}
  0 &= 2 i \mu \eta j^2 \wedge j_1 - e^{- 2 A} \mathd \left[ e^{4 A} (1 -
  \eta^2) \right] \wedge j^2 \wedge \theta + (1 + \eta)^2 e^{4 A} G_3^{\star}
  \wedge \gamma^{\star} - (1 - \eta)^2 e^{4 A} G_3 \wedge \beta \nonumber\\
  &\hspace{1cm} + (1 + \eta) \mathcal{I}_3 \wedge \gamma^{\star} + (1 - \eta)
  \mathcal{I}_3 \wedge \beta \;, 
\end{align}
where we use
\begin{equation}
  2 (1 - \eta^2) j^3 = 3 (1 + \eta) i \gamma \wedge \gamma^{\star} + 3 (1 -
  \eta) i \beta \wedge \beta^{\star} \;  \;, \;  \; \omega \wedge
  \omega^{\star} \wedge j = 2 \eta j^2 \wedge j_1 \; .
\end{equation}
To simplify the above expression, we employ (\ref{six1sep}), written in the
form:
\begin{equation}
  0 = - \frac{2}{3} i \mu j^3 + \frac{1}{2} e^{- 2 A} \mathd e^{4 A} \wedge
  j^2 \wedge \theta - \frac{1}{2} e^{4 A}  \left( \beta \wedge G_3 -
  \gamma^{\star} \wedge G_3^{\star} \right) + \frac{1}{2} e^{4 A} 
  (\mathcal{J}_6 -\mathcal{K}_6) \;, \label{sixformsimp}
\end{equation}
as well as (\ref{sixformalt}). We find
\begin{equation}
  0 = (1 + \eta) \mathcal{I}_3 \wedge \gamma^{\star} + (1 - \eta)
  \mathcal{I}_3 \wedge \beta - 4 \eta e^{2 A} \tmop{Re} \mathcal{J}_5 \wedge
  \theta + (1 - \eta^2) e^{4 A}  (\mathcal{J}_6 -\mathcal{K}_6) \; . 
  \label{JmKeqn}
\end{equation}

Equations (\ref{ibetagammaeqn1}, \ref{ibetagammaeqn2}, \ref{J5eqn},
\ref{ibetagammaeqn3}, \ref{JmKeqn}) constitute nine conditions on the eight
variables $\mathcal{I}_3 \wedge \beta$, $\mathcal{I}_3 \wedge \gamma$,
$\mathcal{J}_4 \wedge \omega$, $\mathcal{J}_4 \wedge \omega^{\star}$,
$\mathcal{J}_6$, $\mathcal{K}_6$, $\mathcal{J}_5 \wedge \theta$, and
$\mathcal{J}_5 \wedge \bar{\theta}$. Thus, one might expect that we can solve
for all eight variables. Indeed this can be done, even without (\ref{JmKeqn});
it is straightforward to check that all of them must vanish:
\begin{equation}
  \mathcal{I}_3 \wedge \beta \; = \; \mathcal{I}_3 \wedge \gamma \; = \; 0
  \;\;, \;\; \mathcal{J}_4 \wedge \omega \; = \;
  \mathcal{J}_4 \wedge \omega^{\star} \; = \; 0 \;\;, \;\;
  \mathcal{J}_6 \; = \; \mathcal{K}_6 \; = \; 0 \;\;, \;\;
  \mathcal{J}_5 \wedge \theta \; = \; \mathcal{J}_5 \wedge \bar{\theta} \; =
  \; 0 \; . \label{furthercons}
\end{equation}
Together with the constraints derived previously, this implies that
$\mathcal{I}_3$, $\mathcal{J}_4$, $\mathcal{I}_5$ and $\mathcal{J}_5$ take the
form:
\begin{equation}
  \mathcal{I}_3 =\mathcal{I}_{\theta \bar{\theta} ; 1, 0} +\mathcal{I}_{; 2,
  1} + c.c. \;  \;, \;  \; \mathcal{J}_4 =\mathcal{J}_{\theta ; 2, 1}
  +\mathcal{J}_{\theta ; 1, 2} \;  \;, \;  \; \mathcal{I}_5
  =\mathcal{I}_{\theta \bar{\theta} ; 2, 1} + c.c. \;  \;, \;  \;
  \mathcal{J}_5 =\mathcal{J}_{\theta \bar{\theta} ; 2, 1} +\mathcal{J}_{\theta
  \bar{\theta} ; 1, 2} \; .
\end{equation}

To extract the relevant components of (\ref{betasep}, \ref{gammasep}), we
wedge them into $\bar{\theta}$ and $\theta$ respectively and apply
(\ref{omegajsep}) to obtain:
\begin{align}
  0 &= - e^{- 4 A} \mathd \left[ e^{4 A} (1 + \eta) \right] \wedge \omega
  \wedge \tilde{j} - e^{2 A}  \left[ j + \tilde{j} \right] \wedge G_3^{\star}
  + \frac{i}{\tau_2} \mathd \tau^{\star} \wedge \omega^{\star} \wedge
  \tilde{j} \nonumber \\
  &\hspace{1cm}+ i\mathcal{J}_4 \wedge \bar{\theta} - e^{2 A}  (1 + \eta)
  \mathcal{I}_5 \;, \label{wedgethetabar} \\
  0 &= - e^{- 4 A} \mathd \left[ e^{4 A} (1 - \eta) \right] \wedge \omega
  \wedge \tilde{j} - e^{2 A}  \left[ j - \tilde{j} \right] \wedge G_3^{\star}
  + \frac{i}{\tau_2} \mathd \tau^{\star} \wedge \omega^{\star} \wedge
  \tilde{j} \nonumber \\
  &\hspace{1cm}+ i\mathcal{J}_4^{\star} \wedge \theta - e^{2 A}  (1 - \eta)
  \mathcal{I}_5 \; . \label{wedgetheta}
\end{align}
Similarly, to extract the relevant components of (\ref{omegasep}), we wedge it
into $j$ and $j_1$ and simplify using (\ref{omegajsep}) to obtain:
\begin{align}
  0 &= - e^{- 2 A} \mathd \left[ e^{4 A} \eta \right] \wedge \omega \wedge
  \tilde{j} - i e^{4 A} \star G_3^{\star} \wedge j +\mathcal{I}_3 \wedge j -
  \eta e^{4 A} \mathcal{I}_5 \;,  \label{wedgej}\\
  0 &= - e^{- 6 A} \mathd \left[ (1 - \eta^2) e^{8 A} \right] \wedge \omega
  \wedge \tilde{j} - e^{4 A} G_3^{\star} \wedge \left[ j - \eta \tilde{j}
  \right] - i e^{4 A} \star G_3^{\star} \wedge j_1 \nonumber\\
  &\hspace{1cm} +\mathcal{I}_3 \wedge j_1 - e^{4 A}  (1 - \eta^2) \mathcal{I}_5 \; . 
  \label{wedgej1}
\end{align}
To simplify these expressions further, we use the identities:
\begin{equation}
  j \wedge \star \hat{\Omega} = - i \tilde{j} \wedge \left( \hat{\Omega}_{2,
  1} - \hat{\Omega}_{1, 2} \right) \;  \;, \;  \; \tilde{j} \wedge \star
  \hat{\Omega} = - i j \wedge \left( \hat{\Omega}_{2, 1} - \hat{\Omega}_{1, 2}
  \right) \;, \label{wedgejidents}
\end{equation}
where the Hodge decomposition is with respect to $\beta$ (or, equivalently,
$\gamma$) and $\hat{\Omega}$ is any three-form satisfying
$\hat{\Omega}_{\theta ; (1, 1)^{\tmop{NP}}} = \hat{\Omega}_{\bar{\theta} ; (1,
1)^{\tmop{NP}}} = 0$. To prove these identities, note that we can decompose
$\hat{\Omega} = j \wedge v + \tilde{j} \wedge w + \ldots$, where the omitted
terms vanish when wedged into $j$ and $\tilde{j}$. One can then show using the
primitivity decomposition that
\begin{equation}
  \star \hat{\Omega} = - i \left( v_{1, 0} - v_{0, 1} \right) \wedge \tilde{j}
  - i (w_{1, 0} - w_{0, 1}) \wedge j + \ldots \; .
\end{equation}
The identities (\ref{wedgejidents}) are now easily verified.

Thus, (\ref{wedgej}, \ref{wedgej1}) become:{\footnote{To clarify notation,
$G_{(p, q)}^{\star} \equiv \left[ G_{(p, q)} \right]^{\star} = \left[
G^{\star} \right]_{(q, p)} \neq \left[ G^{\star} \right]_{(p, q)}$.}}
\begin{align}
  0 &= - e^{- 2 A} \mathd [e^{4 A} \eta] \wedge \omega \wedge
  \tilde{j} - e^{4 A} G_{(1, 2)}^{\star} \wedge \tilde{j} +\mathcal{I}_{(2,
  1)} \wedge j - \eta e^{4 A} \mathcal{I}_{(3, 2)} \;, \\
  0 &= e^{4 A} G_{(2, 1)}^{\star} \wedge \tilde{j} +\mathcal{I}_{(1, 2)}
  \wedge j - \eta e^{4 A} \mathcal{I}_{(2, 3)} \;,  \label{jtwG21}\\
  0 &= - e^{- 6 A} \mathd [(1 - \eta^2) e^{8 A}] \wedge \omega
  \wedge \tilde{j} - 2 e^{4 A} G_{(1, 2)}^{\star} \wedge j_2 +\mathcal{I}_{(2, 1)} \wedge j_1 - e^{4 A}  (1 - \eta^2)
  \mathcal{I}_{(3, 2)} \;,  \label{j2G12}\\
  0 &= \mathcal{I}_{(1, 2)} \wedge j_1 - e^{4 A}  (1 - \eta^2)
  \mathcal{I}_{(2, 3)} \;, 
\end{align}
where $j_1 = \tilde{j} - \eta j$ and $j_2 = j - \eta \tilde{j}$. Similarly, (\ref{wedgethetabar},
\ref{wedgetheta}) become:
\begin{align}
  0 &= - e^{- 4 A} \mathd [e^{4 A} (1 + \eta)] \wedge \omega
  \wedge \tilde{j} - e^{2 A} [ j + \tilde{j}] \wedge G_{(1,
  2)}^{\star} + i\mathcal{J}_{(3, 1)} \wedge \bar{\theta} - e^{2 A}  (1 +
  \eta) \mathcal{I}_{(3, 2)} \;, \\
  0 &= - e^{2 A} [ j + \tilde{j}] \wedge G_{(2, 1)}^{\star} +
  \frac{i}{\tau_2} \mathd \tau^{\star} \wedge \omega^{\star} \wedge \tilde{j}
  + i\mathcal{J}_{(2, 2)} \wedge \bar{\theta} - e^{2 A}  (1 + \eta)
  \mathcal{I}_{(2, 3)} \;,  \label{jpljtwG21}\\
  0 &= - e^{- 4 A} \mathd [e^{4 A} (1 - \eta)] \wedge \omega
  \wedge \tilde{j} - e^{2 A} [ j - \tilde{j}] \wedge G_{(1,2)}^{\star} + i\mathcal{J}_{(2, 2)}^{\star} \wedge \theta
   - e^{2 A}  (1 - \eta) \mathcal{I}_{(3, 2)} \;, \\
  0 &= - e^{2 A} [ j - \tilde{j}] \wedge G_{(2, 1)}^{\star} +
  \frac{i}{\tau_2} \mathd \tau^{\star} \wedge \omega^{\star} \wedge \tilde{j}
  + i\mathcal{J}_{(3, 1)}^{\star} \wedge \theta - e^{2 A}  (1 - \eta) \mathcal{I}_{(2, 3)} \; .  \label{jmjtwG21}
\end{align}
We combine these equations to eliminate $G_3$ and $\mathd \tau$, leaving:
\begin{align}
  0 &= 2\mathcal{I}_{(1, 2)} \wedge j + i e^{2 A} \mathcal{J}_{(2, 2)}
  \wedge \bar{\theta} - i e^{2 A} \mathcal{J}_{(3, 1)}^{\star} \wedge \theta -
  4 \eta e^{4 A} \mathcal{I}_{(2, 3)} \;,  \label{cond1}\\
  0 &= \mathcal{I}_{(1, 2)} \wedge j_1 - e^{4 A}  (1 - \eta^2)
  \mathcal{I}_{(2, 3)} \;,  \label{cond2}\\
  0 &= - 2\mathcal{I}_{(2, 1)} \wedge j + i e^{2 A} \mathcal{J}_{(3, 1)}
  \wedge \bar{\theta} - i e^{2 A} \mathcal{J}_{(2, 2)}^{\star} \wedge \theta
  \;,  \label{cond3}\\
  0 &= -\mathcal{I}_{(2, 1)} \wedge \left[ j_1 + 2 \eta j \right] + i e^{2
  A} \mathcal{J}_{(3, 1)} \wedge \bar{\theta} + i e^{2 A} \mathcal{J}_{(2,
  2)}^{\star} \wedge \theta - e^{4 A}  (1 - \eta^2) \mathcal{I}_{(3, 2)} \; . 
  \label{cond4}
\end{align}

Applying the Hodge decomposition to (\ref{tausep}, \ref{jjsep}) and extracting
the relevant components, we obtain:
\begin{align}
  0 &= \mathcal{J}_{(3, 2)} \;, \\
  0 &= - \frac{i}{\tau_2}  \bar{\partial}_{\Pi} \tau \wedge \tilde{j}
  \wedge j_2 + \frac{1}{4} e^{2 A}  (1 - \eta^2) \omega^{\star} \wedge
  \hat{G}_{(2, 1)} + \frac{1}{2}  (1 - \eta^2) \mathcal{J}_{(2, 3)} \;, \\
  0 &= \frac{1}{2} e^{- 8 A}  \bar{\partial}_{\Pi} [(1 - \eta^2) e^{8 A}] \wedge \tilde{j} \wedge j_2 - \frac{1}{4} e^{2 A}  (1 - \eta^2)
  \omega^{\star} \wedge \hat{G}_{(1, 2)}^{\star} - (1 - \eta^2) [ \tmop{Re} \mathcal{J}_5 ]_{(2, 3)} \;,
\end{align}
where $\bar{\partial}_{\Pi}$ is the projection of the scalar gradient onto
antiholomorphic directions along the base, $j_2 = j - \eta \tilde{j}$, and
$\hat{G}_3 = G_{\theta \bar{\theta} ; 1, 0} + G_{\theta \bar{\theta} ; 0, 1} +
G_{; 2, 1} + G_{; 1, 2}$ consists of the components of $G_3$ with an even
number of legs along the fiber. The latter two equations can be usefully
restated using the identity:
\begin{equation}
  v_{1, 0} \wedge J_2 = \frac{1}{2} w_{0, 1} \wedge \Omega_2 \;  \;
  \leftrightarrow \;  \; w_{0, 1} \wedge J_2 = - \frac{1}{2} v_{1, 0} \wedge
  \bar{\Omega}_2 \;,
\end{equation}
for any $\tmop{SU}(2)$ structure, where $v$ and $w$ point along the base. In
particular,
\begin{equation}
  \hat{v}_{2, 1} \wedge j_2 = \frac{1}{4}  (1 - \eta^2)  \hat{w}_{1, 2} \wedge
  \omega \;  \; \leftrightarrow \;  \; \hat{w}_{1, 2} \wedge j_2 = -
  \hat{v}_{2, 1} \wedge \omega^{\star} \;,
\end{equation}
for any $v$, $w$ with an even number of legs along the fiber, since $j_2 =
s_{\varphi}^2 J_2$. Thus, taking $\mathcal{J}_5 = \left[ \frac{1}{2} \eta j^2
+ j \wedge j_1 \right] \wedge \mathcal{J}_1$, where $\mathcal{J}_1$ points
along the base, we obtain:
\begin{align}
  \mathcal{J}_{(1, 0)} &= 0 \;,  \label{cond5}\\
  0 &= \frac{1}{2} \mathcal{J}_{(0, 1)} \wedge \omega \wedge \tilde{j} -
  \frac{i}{\tau_2} \mathd \tau \wedge \omega \wedge \tilde{j} - e^{2 A} j_2
  \wedge \hat{G}_{(2, 1)} \;, \\
  0 &= \frac{1}{2} e^{- 8 A} \mathd \left[ (1 - \eta^2) e^{8 A} \right]
  \wedge \omega \wedge \tilde{j} - \left[ \tmop{Re} \mathcal{J}_1 \right]_{(0,
  1)} \wedge \omega \wedge \tilde{j} + e^{2 A} j_2 \wedge \hat{G}_{(1,
  2)}^{\star} \; . 
\end{align}
Combining with (\ref{jtwG21}, \ref{j2G12}, \ref{jpljtwG21}) to eliminate
$G_3$, we find:
\begin{align}
  0 &= \mathcal{I}_{(2, 1)} \wedge j_1 - e^{4 A}  (1 - \eta^2)
  \mathcal{I}_{(3, 2)} - 2 e^{2 A}  \left[ \tmop{Re} \mathcal{J}_1
  \right]_{(0, 1)} \wedge \omega \wedge \tilde{j} \;,  \label{cond6}\\
  0 &= e^{2 A} \mathcal{J}_{(0, 1)}^{\star} \wedge \omega^{\star} \wedge
  \tilde{j} - 2 i e^{2 A} \mathcal{J}_{(2, 2)} \wedge \bar{\theta} + 2 e^{4 A}
  (1 + \eta)^2 \mathcal{I}_{(2, 3)} - 2 (1 + \eta) \mathcal{I}_{(1, 2)}
  \wedge j \; .  \label{cond7}
\end{align}

The conditions (\ref{cond1} -- \ref{cond4}, \ref{cond5}, \ref{cond6},
\ref{cond7}) constitute seven equations in seven unknowns:
$\mathcal{I}_{\left( 2, 1 \right)} \wedge j$, $\mathcal{I}_{\left( 2, 1
\right)} \wedge \tilde{j}$, $\mathcal{I}_{\left( 3, 2 \right)}$,
$\mathcal{J}_{\left( 3, 1 \right)} \wedge \bar{\theta}$, $\mathcal{J}_{\left(
2, 2 \right)} \wedge \bar{\theta}$, $\mathcal{J}_{\left( 1, 0 \right)} \wedge
\omega^{\star} \wedge \tilde{j}$ and $\mathcal{J}_{(0, 1)} \wedge \omega
\wedge \tilde{j}$. One can check that the only solution is
\begin{equation}
  \mathcal{I}_{(2, 1)} \wedge j =\mathcal{I}_{(2, 1)} \wedge \tilde{j} = 0 \; 
  \;, \;  \; \mathcal{I}_{(3, 2)} = 0 \;  \;, \;  \; \mathcal{J}_{(3, 1)}
  \wedge \bar{\theta} =\mathcal{J}_{(2, 2)} \wedge \bar{\theta} = 0 \;  \;, \;
  \; \mathcal{J}_{(1, 0)} =\mathcal{J}_{(0, 1)} = 0 \; .
\end{equation}
Taken together with the constraints derived previously, we see that all
separation forms must vanish, so that the supersymmetry conditions are
manifestly $\tmop{SL} (2, \mathbb{R})$ covariant.

\bibliographystyle{JHEP}
\renewcommand{\refname}{References}
\addcontentsline{toc}{section}{References}
\bibliography{paperdraft}

\end{document}